\documentclass[prd,showpacs,groupedaddress,superscriptaddress,nofootinbib,floatfix,preprintnumbers,tightenlines,11pt,longbibliography]{revtex4}
\usepackage[paperwidth=210mm,paperheight=297mm,centering,hmargin=2.4cm,vmargin=2.4cm]{geometry}

\usepackage{float}
\usepackage{nicefrac}
\usepackage{slashed}
\usepackage{mathtools}
\usepackage{amsfonts} 
\usepackage{amssymb} 
\usepackage{amsmath} 
\usepackage{graphicx} 
\usepackage{subfigure} 
\usepackage{array} 
\usepackage{dcolumn} 
\usepackage{bm} 
\usepackage{latexsym} 
\usepackage{longtable} 
\usepackage{hyperref} 
\usepackage{verbatim}
\usepackage{epsfig}
\usepackage{color}
\usepackage{xcolor}
\usepackage{cancel}
\usepackage{braket}
\newcommand{\nn}[0]{\nonumber}

\begin{document}

\title{
Long-range electroweak amplitudes of single hadrons \\
from Euclidean finite-volume correlation functions
}

\author{Ra\'ul A.~Brice\~no}
\email[]{rbriceno@jlab.org}
\affiliation{Thomas Jefferson National Accelerator Facility, Newport News, Virginia 23606, USA}
\affiliation{ Department of Physics, Old Dominion University, Norfolk, Virginia 23529, USA}
\author{Zohreh Davoudi}
\email[]{davoudi@umd.edu}
\affiliation{Maryland Center for Fundamental Physics and Department of Physics,University of Maryland, College Park, Maryland 20742, USA
}
\affiliation{RIKEN Center for Accelerator-based Sciences, Wako 351-0198, Japan
}
\author{Maxwell T. Hansen}
\email[]{maxwell.hansen@cern.ch}
\affiliation{Theoretical Physics Department, CERN, 1211 Geneva 23, Switzerland}

\author{Matthias R.~Schindler}
\email[]{mschindl@mailbox.sc.edu}
\affiliation{Department of Physics and Astronomy, University of South Carolina, Columbia, South Carolina 29208, USA}
\author{Alessandro Baroni}
\email[]{abaro008@odu.edu}
\affiliation{Theoretical Division, Los Alamos National Laboratory, Los Alamos, New Mexico 87545, USA}
\affiliation{Department of Physics and Astronomy, University of South Carolina, Columbia, South Carolina 29208, USA}
%
\date{\today}

 
\preprint{JLAB-THY-19-3051}
\preprint{CERN-TH-2019-189}
\preprint{UMD-PP-019-06}
\begin{abstract}
A relation is presented between single-hadron long-range matrix elements defined in a finite Euclidean spacetime, and the corresponding infinite-volume Minkowski amplitudes. This relation is valid in the kinematic region where any number of two-hadron states can simultaneously go on shell, so that the effects of strongly-coupled intermediate channels are included. These channels can consist of non-identical particles with arbitrary intrinsic spins. The result accommodates general Lorentz structures as well as non-zero momentum transfer for the two external currents inserted between the single-hadron states. The formalism, therefore, generalizes the work by Christ et al.~[Phys.Rev. D91 114510 (2015)], and extends the reach of lattice quantum chromodynamics (QCD) to a wide class of new observables beyond meson mixing and rare decays. Applications include Compton scattering of the pion ($\pi \gamma^\star \to [\pi \pi, K \overline K] \to \pi \gamma^\star$), kaon ($K \gamma^\star \to [\pi K, \eta K] \to K \gamma^\star$) and nucleon ($N \gamma^\star \to N \pi  \to N \gamma^\star$), as well as double-$\beta$ decays, and radiative corrections to the single-$\beta$ decay, of QCD-stable hadrons. The framework presented will further facilitate generalization of the result to studies of nuclear amplitudes involving two currents from lattice QCD.
\end{abstract}
\maketitle
\nopagebreak

\section{Introduction \label{sec:intro}}
\noindent
Long-range electroweak matrix elements play a central role in modern hadronic physics, from precision tests of the Standard Model (SM) to investigations into the inner structure of strongly-interacting particles. In this work, a subset of such matrix elements are considered, defined with a single incoming and a single outgoing hadron, coupled to two local currents that are displaced in time. Key examples of processes for which these types of matrix elements are needed include Compton scattering, double-$\beta$ decays, radiative corrections to single-$\beta$ decays, $K$-$\overline{K}$ oscillations and rare meson decays:

\begin{itemize}
\item Deeply virtual Compton scattering, i.e.~the conversion from a virtual photon to a real photon via scattering off a charged hadron ($\gamma^*h\to \gamma h$), allows one to extract the generalized parton distributions of the target, as proposed by Ji~\cite{Ji:1996nm}. Such a process can be studied at leading order in quantum electrodynamics (QED) by performing appropriate Fourier transforms of matrix elements involving time-displaced electromagnetic currents. 

\item  A large experimental effort is dedicated to searches for a lepton-number-violating process in nature, namely the neutrinoless double-$\beta$ decay of certain nuclear isotopes, e.g., see Refs.~\cite{DellOro:2016tmg,Dolinski:2019nrj,Henning:2016fad,Rodejohann:2011mu}. 
 If observed, such a decay would establish that neutrinos are Majorana fermions. In order to understand the implication of a potential observation for extensions of the Standard Model, reliable theoretical constraints on the relevant QCD matrix elements are necessary. Moreover, as a supplement to the direct extraction of nucleon matrix elements, nuclear effective field theories~\cite{Prezeau:2003xn,Menendez:2011qq,Cirigliano:2017tvr,Cirigliano:2018hja,Cirigliano:2018yza} indicate that the $\pi^+ \to \pi^-$ inversion, as well as short distance nuclear effects, may be important contributions to the $nn \to ppee$ conversion process. In fact, calculations of the matrix elements relevant for these processes have already started~\cite{Nicholson:2018mwc,Shanahan:2017bgi,Tiburzi:2017iux,Tuo:2019bue}, and a class of these computations require bi-local matrix elements~\cite{Shanahan:2017bgi,Tiburzi:2017iux,Feng:2018pdq,Detmold:2018zan,Tuo:2019bue}. 

\item Interest in constraining radiative corrections to nuclear $\beta$ decays has grown in recent years in light of a discrepancy among theoretical determinations~\cite{Marciano:2005ec,Seng:2018yzq,Czarnecki:2019mwq}, leading to a $\sim 2-3\sigma$ deviation in the CKM unitary test from $V_{ud}$. As an indirect input to calibrate several contributions to the $\beta$-decay calculations, constraining the time-ordered product of a weak and an electromagnetic current between a neutron and a proton state, for a number of fixed momentum transfers, would be immensely valuable~\cite{Cirigliano:2019jig}.

\item In the strange and charmed mesonic sector, bi-local matrix elements can be used to compute the mass splittings between neutral mesons. Another set of observables based on this class of matrix elements arise in rare decays of such mesons, including $K \to \pi l^+l^-$ and $K \to \pi \nu \bar{\nu}$, relevant to searches of CP violation. These processes are currently being measured, for example, in the NA62 experiment at CERN.%
\footnote{Further examples can be found, e.g., in a series of recent USQCD whitepapers~\cite{Lehner:2019wvv, Cirigliano:2019jig, Detmold:2019ghl, Kronfeld:2019nfb}.}
\end{itemize}

At present, the most reliable approach to determine such matrix elements from the underlying theory is lattice QCD, a numerical method for statistically estimating QCD correlation functions using Monte Carlo importance sampling. However, since lattice-QCD calculations are necessarily performed in a finite Euclidean spacetime,\footnote{For exploratory studies of real-time dynamics in simple theories with classical computations based on Monte Carlo methods, and with quantum computations based on direct implementation of the Hamiltonian time evolution, see Refs.~\cite{pichler2016real,Alexandru:2016gsd, Martinez:2016yna, Klco:2018kyo, Davoudi:2019bhy}. The generalization of these methods to QCD remains formally and practically challenging.} 
 the relation between the calculated quantity and the physical observable is not always straightforward. 

 The formalism for extracting long-range matrix elements from finite-volume Euclidean correlation functions was first derived in Refs.~\cite{Christ:2010gi, Christ:2014qaa, Christ:2015pwa}, in order to determine the long-range contributions to the $K_L$-$K_S$ mass splitting and the CP-violating parameter $\epsilon_K$. This framework has enabled determinations of the mass splitting of neutral kaons via lattice QCD~\cite{Christ:2012se, Bai:2014cva, Christ:2014qwa, Bai:2018mdv}. Long-range matrix elements occurring in rare decays of the kaon, such as in $K \to \pi l^+l^-$ and $K \to \pi \nu \bar{\nu}$ processes~\cite{Christ:2016mmq, Bai:2017fkh, Bai:2018hqu,Christ:2019dxu}, and hadronic double-$\beta$ decays~\cite{Shanahan:2017bgi,Tiburzi:2017iux,Feng:2018pdq,Detmold:2018zan}, have also been studied in recent years with lattice QCD. 
 
In this paper, we generalize the previous work to incorporate matrix elements with multiple two-particle intermediate states propagating between the two currents. The result presented holds for kinematics for which any number of two-hadron states can go on shell. We further accommodate particles with intrinsic spin as well as channels with non-identical particles. In addition, the two local currents in this work are allowed to have a generic structure, including any number of  Lorentz indices and non-zero energy and momentum injection. Finally, the expressions presented account for the unphysical mixing of different angular momentum states due to the reduced symmetry of a cubic volume, as well as the physical mixing of different orbital angular momenta in systems with nonzero spin. As a nontrivial check, given these general considerations, the result obtained is proved to satisfy the unitarity of the physical amplitude to all orders. The general form presented is an essential step towards extensions to multi-hadron bi-local matrix elements.
 
 In order to explore some of the key ideas of Refs.~\cite{Christ:2010gi, Christ:2014qaa, Christ:2015pwa}, as well as the general framework of the present study, it is useful to begin with the spectral decomposition of the Euclidean finite-volume matrix element
\begin{align}
\int d^3  \boldsymbol x \, e^{- i \boldsymbol q \cdot \boldsymbol x} \, \langle M,L|  {J}_E(\tau, \boldsymbol x) J(0)|M,L\rangle
= L^3 \sum_{n}
e^{-\tau (E_n-M)}
\langle M,L|  {J}(0) | P_n,L \rangle \langle   P_n,L| J(0)|M,L\rangle \,,
\label{eq:example}
\end{align}
where $J(0)$ is a generic local current and $J_E(\tau, \boldsymbol x) \equiv e^{H\tau}J(0, \boldsymbol x)e^{-H\tau}$ defines its Euclidean time translation. In writing Eq.~\eqref{eq:example}, two important assumptions are made: First, we neglect the effects of the finite temporal direction, focusing only on the spatial volume with periodicity $L$. This is well motivated as lattice-QCD calculations typically work with $T > L$ such that the finite-$T$ effect is a sub-leading uncertainty.  Second, we have introduced a finite-volume single-hadron state, denoted $\vert M, L \rangle$, and formally defined by the $L$-dependence of its eigenvalue, $E(L)$: $\lim_{L \to \infty}E(L) = M$, where $M$ is the physical mass.\footnote{For a fixed value of spatial momentum, as well as fixed internal quantum numbers, this defines a unique state. In theories without a mass gap, e.g.~QCD+QED, this separation fails and the distinction is less straightforward.} For the purpose of the introduction only, the state is assumed to have a vanishing spatial momentum. The introductory example is further simplified by taking the two currents to be the same and to be Hermitian. These restrictions will be removed in arriving at the main result of this work. Finally, it should be stressed that the complete set of states inserted on the right-hand side of Eq.~\eqref{eq:example}  is a proper sum, since the finite-volume boundary conditions lead to a discrete spectrum. The sum runs over an infinite tower of states with the same quantum numbers as $J(0) \vert M, L \rangle$. Due to the Fourier transform in Eq.~\eqref{eq:example}, the inserted states are understood to be projected to a definite spatial momentum, $ \bm{P}_n = \boldsymbol q$.
 
The corresponding infinite-volume Minkowski observable that we aim to determine is given by
 \begin{equation}
 \label{eq:calTintro}
\mathcal T(\omega, \textbf q) = - \lim_{\epsilon \to 0^+} \lim_{L \to \infty}   2 M L^6 \sum_{n}
\frac{
\vert \langle   P_n,L| J(0)|M,L\rangle \vert^2}{ \omega - [E_n(L)-M] + i \epsilon } \,,
 \end{equation}
 where $2 M L^3$ accounts for the normalization of $ \vert M, L \rangle$ and an extra factor of $L^3$ results from performing the spatial Fourier transform.\footnote{In fact we will use a different, but equivalent form for the infinite-volume amplitude in the following sections. We present this version only because it gives additional intuition on the relation between the finite- and infinite-volume amplitudes. For detailed discussion on the form of $\mathcal T$ given here, and prospects for directly evaluating Eq.~\eqref{eq:calTintro} numerically, see Refs.~\cite{Hansen:2017mnd,Bulava:2019kbi}.} Thus, the task at hand is to relate Eqs.~\eqref{eq:example} and \eqref{eq:calTintro}. The basic approach for achieving this can be understood as a two-step procedure. First, one must replace the $\tau$-dependent exponentials ($e^{- E_n \tau}$) with poles ($1/(\omega - E_n)$). In the case that $\omega + M < E$, this is achieved by applying the integral operator $\int_0^\infty d \tau e^{\omega \tau}$. However, as stressed in Refs.~\cite{Christ:2010gi, Christ:2014qaa, Christ:2015pwa}, many interesting cases arise where the low-lying finite-volume states are below the target $\omega$ value, so that the indicated integral is exponentially divergent. 
 
Various approaches have been considered to treat this problem~\cite{Christ:2010gi, Christ:2014qaa, Christ:2015pwa, Christ:2016mmq, Bai:2017fkh, Bai:2018hqu}. In the present work, we are particularly inspired by the techniques that have been applied for the hadronic vacuum polarization (HVP) contribution to the anomalous magnetic moment of the muon~\cite{Bernecker:2011gh,Meyer:2011um,Francis:2013qna,DellaMorte:2017dyu,Blum:2018mom,Meyer:2018til,Gerardin:2019rua}, as well as the incredible success in extracting excited finite-volume states to determine hadronic scattering amplitudes~\cite{Briceno:2017max}. Following these ideas, one can envision solving the generalized eigenvalue problem (GEVP) on a matrix of lattice-QCD correlators, in order to reliably determine as many low-lying energies, $E_n(L)$, and matrix elements, $ \vert \langle   P_n,L| J(0)|M,L\rangle \vert^2$, as possible. These can then be subtracted from Eq.~(\ref{eq:example}), rendering the $\int_0^\infty d \tau e^{\omega \tau}$ integral convergent. In a second step, the subtracted terms are added back in after integration, but with the exponential time dependence replaced by a pole, as detailed in Sec.~\ref{sec:main_result}. The result of this construction is an intermediate quantity, denoted by $ T_L(\omega, \boldsymbol q)$:
\begin{equation}
 T_L(\omega, \boldsymbol q) \equiv   - 2 M L^6 \sum_{n}
\frac{
\vert \langle   P_n,L| J(0)|M,L\rangle \vert^2}{ \omega - [E_n(L)-M]   }  \,.
 \end{equation}
 
These manipulations address the effects of the Euclidean signature, but not yet those arising from the finite volume. The remaining task is then to determine the correction term $\Delta T_L(\omega, \boldsymbol q) \equiv \mathcal T(\omega, \boldsymbol q) -  T_L(\omega, \boldsymbol q)$, to be added to the sum over finite-volume poles to reach the final physical result. As a special case, it is instructive to consider $\Delta T_L(\omega, \boldsymbol{q})$ when the value of $\omega +M$ is taken below all single- and multi-hadron thresholds. In this case, the integral over $e^{(\omega + M - E_n) \tau}$ converges for all $E_n$. It turns out that in this kinematic region, $\Delta T_L = \mathcal{O}(e^{-m L})$, where $m$ is the mass of the lightest degree of freedom, i.e., the pion in QCD. 
   
In Sec.~\ref{sec:main_result}, we provide a closed form for $\Delta T_L(\omega, \boldsymbol q)$ in the case that the center-of-mass energy of the incoming current plus hadron is above any number of two-particle thresholds, but is below the three-hadron production thresholds. This correction can be obtained, up to neglected $\mathcal O(e^{- m L})$ terms, provided that the relevant $2 \to 2$ scattering amplitudes and $1 \to 1$ and $ 1\to 2$ transition amplitudes are determined. This can be achieved through independent, dedicated lattice-QCD calculations based on well-established methods. In particular, by applying L\"uscher-like formalisms~\cite{Luscher:1991n1, Rummukainen:1995vs, Kim:2005gf, He:2005ey, Davoudi:2011md, Hansen:2012tf, Briceno:2012yi, Briceno:2013lba, Briceno:2014oea}, one can determine the infinite-volume scattering amplitudes of general two-hadron states from finite-volume spectra, see Refs.~\cite{Wilson:2015dqa, Briceno:2016mjc, Brett:2018jqw, Guo:2018zss, Andersen:2017una, Andersen:2018mau, Dudek:2014qha, Dudek:2016cru, Woss:2018irj, Woss:2019hse, Orginos:2015aya, Berkowitz:2015eaa, Wagman:2017tmp} for recent examples. This approach has been further generalized to give relations between finite-volume matrix elements and electroweak amplitudes~\cite{Lellouch:2000, Meyer:2011um, Briceno:2012yi, Briceno:2014uqa, Feng:2014gba, Briceno:2015csa, Briceno:2015tza, Baroni:2018iau}.\footnote{For recent reviews on these formal developments and their numerical implementations, we point the readers to Refs.~\cite{Briceno:2017max, Yamazaki:2015nka, Davoudi:2017ddj, Davoudi:2018wgb}.} The upshot is that the required methodology for extracting all the ingredients of $\Delta T_L(\omega, \boldsymbol q)$ is already established.

Finally, the formalism presented can be used to reduce systematic uncertainties of long-range matrix elements of hadrons even below two-particle thresholds, by identifying and removing contributions from intermediate states that cannot go on shell.  A similar approach has been applied to lattice-QCD calculations of the muonic HVP, where the knowledge of the $\pi\pi\to\pi\pi$ and $\gamma^* \to \pi\pi$ amplitudes allows one to estimate the finite-volume effects, as suggested in Refs.~\cite{Bernecker:2011gh,Meyer:2011um} and implemented in Refs.~\cite{Izubuchi:2018tdd, Gerardin:2019rua}.

This paper is organized as follows: Sec.~\ref{sec:main_result} presents the main results, including the framework that enables the determination of $\Delta T_L$. Sec.~\ref{sec:FV} contains the derivation of $\Delta T_L$, relying largely on combining existing ideas in the literature to determine scattering and transition amplitudes in the single- and two-hadron sectors. In Sec.~\ref{sec:check}, a strong check on the formalism is presented showing that the result obtained is consistent with unitarity constraints on the infinite-volume amplitude. In order to demonstrate an application of the formalism, in Sec.~\ref{sec:example} we present an example with a single-channel intermediate state and evaluate numerically all the building blocks for a specific toy example. We conclude in Sec.~\ref{sec:conclusion} with a summary and outlook.

\section{The relation between finite-volume Euclidean and infinite-volume Minkowski amplitudes
\label{sec:main_result}}	
\noindent

The main result of this work can be compactly written as 
\begin{equation}
\mathcal T(\omega,\bm{q}) =  \int_{-\infty}^\infty d \tau \, e^{\omega  \tau }       \left[ G_L(\tau, \bm{q})-G_L^{< N}(\tau, \bm{q}) \right] + 
\left[T^{< N}_L(\omega, \bm{q})+ \Delta    T_L(\omega, \bm{q})\right]_{\mathcal M ,\mathcal H} \ .
\label{eq:master}
\end{equation}
Here, the dependence of the functions on the four-momenta of the initial and final states is left implicit. The left-hand side is the desired infinite-volume Minkowski amplitude, and the right-hand side is a carefully constructed combination of finite-volume Euclidean and Minkowski quantities, which all can be obtained from finite-volume Euclidean correlation functions. In a nutshell, $G_L$ is a finite-volume Euclidean matrix element, $G_L^{< N}$ is a reconstruction of its low-lying states (using e.g., GEVP methods), $T^{< N}_L$ is a corresponding sum of the low-lying finite-volume poles and $\Delta    T_L$ is the correction term to remove finite-volume effects. The aim of this section is to provide the exact definitions of these functions and the relations among them. As a final general comment, one must note that for kinematics where intermediate two-particle states may go on-shell, the last two terms, $T^{< N}_L$ and $\Delta    T_L$,  contain poles that must exactly cancel. For this reason, the two quantities must be treated consistently. This is indicated by the square brackets and the subscript $\mathcal{M},\mathcal{H}$, referring respectively to the relevant $2 \to 2$ and $1 \to 2~(2 \to 1)$ scattering amplitudes. This point will be explained in detail towards the end of this section.

We first introduce the basic kinematic notation used throughout. The three-momentum of the incoming hadron state is denoted by $\bm{P}_i$ and that of the outgoing state by $\bm{P}_f$. The corresponding four-momentum of the initial state is then given by
\begin{equation}
P_i^\mu \equiv (E_i, \bm{P}_i) = \Big (  \sqrt{M_i^2 + \bm{P}_i^2} , \bm{P}_i   \Big ) \,,
\end{equation}
with an analogous relation for the final state given by $i \to f$. Here, $M_i$ and $M_f$ are the physical particle masses. The infinite-volume hadronic states are denoted by $\vert P_i  \rangle$ and $\langle P_f  \vert$ and satisfy the standard relativistic normalization:
\begin{align}
\label{eq:IV_norm}
\langle P\rvert P^\prime\rangle=2E\left(2\pi\right)^3 \delta^3(\bm{P}-\bm{P}^\prime)\, .
\end{align}

Next we introduce two local, Minkowski-signature currents $\mathcal J_A(x)$ and $\mathcal O_B(x)$, with $x^\mu \equiv (t, \bm{x})$. Here, $A$ and $B$ are collective indices that specify the quantum numbers of the currents. They can specify scalar, axial, vector, tensor or other types of currents. The underlying Lorentz structure plays no role in the following discussions. The goal is to relate a finite-volume Euclidean correlation function to the infinite-volume amplitude appearing on the left-hand side of Eq.~(\ref{eq:master}), defined as
\begin{equation}
\label{eq:TAdef}
\mathcal T(\omega,\bm{q}) \equiv   i \int d^4 x \, e^{i\omega t - i\bm{q} \cdot \bm{x}} \, \langle P_f \vert  \,  \text{T} \{ \mathcal J_A(x)  \mathcal O_B(0) \} \, \vert P_i   \rangle_{\text{conn.}} \,,
\end{equation}
where again the $P_i$ and $P_f$ dependence on the left-hand side is implicit. Here T denotes the standard time ordering\footnote{Not to be confused with the temporal length of spacetime that was introduced earlier.} and the subscript ``conn.'' indicates that only the connected contributions to the matrix element are considered. This distinction is only relevant in the forward limit when $P_i = P_f$.

In a finite cubic volume, with periodicity $L$ in each of the three spatial directions, the $L$-dependent shifts to the masses, $M_i$ and $M_f$, are exponentially suppressed, scaling as $e^{- m L}$ where $m$ is the mass of the lightest low-energy degree of freedom in the theory. Here we assume $mL\gg 1$, such that these corrections can be neglected. Thus the four-vectors $P_i$ and $P_f$ also label the finite-volume states, denoted by $\vert P_i, L \rangle$ for the incoming hadron and $\langle P_f, L \vert$ for the outgoing. In the convention of this work, finite-volume states are normalized to unity,
\begin{align}
\label{eq:FV_norm}
\langle P , L\rvert P^\prime , L\rangle&= \delta_{\bm{P},\bm{P}^\prime}\,,
\end{align}
keeping in mind that in a periodic cubic volume three-momenta satisfy: $\bm{P}=2\pi \bm{n}/L$ and $\bm{P}^\prime=2\pi\bm{n}^\prime/L$, with $\bm{n},\bm{n}^\prime\in\mathbb{Z}^3$. 

Introducing $\mathcal J^E_A(\tau, \bm{x})$ and $ \mathcal O^E_B(\tau, \bm{x})$ as Euclidean counterparts of the local currents in Eq.~(\ref{eq:TAdef}),
\begin{equation}
 \mathcal J_A^E(\tau , \bm{x}) \equiv e^{H \tau}  \mathcal J_A(0, \bm{x}) e^{- H \tau} \,, \qquad   \mathcal O_B^E(\tau, \bm{x}) \equiv e^{ H \tau}  \mathcal O_B(0, \bm{x}) e^{- H \tau}  \,,
\end{equation}
 one can define the lattice-QCD correlator most closely related to $\mathcal T$ as
\begin{equation}
\label{eq:GFV}
G_L(\tau, \bm{q}) \equiv 2 L^3 \sqrt{E_i  E_f } \int_L d^3 \bm{x}  \, e^{- i  \bm{q} \cdot \bm{x}} \, \langle P_f, L \vert  \,  \text{T}_{\!E} \{ \mathcal J^E_A(\tau, \bm{x})  \mathcal O_B^E(0) \} \, \vert P_i, L   \rangle \,.
\end{equation}
In the forward limit, the disconnected contribution
\begin{equation}
G_L^{\text{disc}}(\tau, \bm{q})   \equiv  2 L^3 \sqrt{E_i  E_f }  \langle P_f, L \vert  P_i, L   \rangle \int_L d^3 \bm{x}  \, e^{- i  \bm{q} \cdot \bm{x}} \, \langle 0 \vert     \text{T}_{\!E} \{ \mathcal J^E_A(\tau, \bm{x})  \mathcal O_B^E(0) \}   \vert 0   \rangle  \,,
\label{eq:Gdis}
\end{equation}
must be subtracted from Eq.~(\ref{eq:GFV}) in order to obtain the purely connected piece of the matrix element as required by Eq.~(\ref{eq:TAdef}). For notational brevity, the hadron momentum labels $P_i$ and $P_f$ are also omitted from the arguments of the correlation functions. 

To understand the issues in extracting $\mathcal T$ from $  G$, it is instructive to first perform a spectral decomposition of the latter. Defining
\begin{align}
\label{eq:c}
c_n & \equiv   
2 L^3 \sqrt{E_i  E_f } \int_L d^3 \bm{x}  \, e^{- i  \bm{q} \cdot \bm{x}} \,  \langle P_f, L \vert   \mathcal J_A(0, \bm{x}) \vert n, L \rangle \langle n, L \vert  \mathcal O_B(0)    \vert P_i, L   \rangle \,, \\
\bar c_{\bar n} & \equiv   
2 L^3 \sqrt{E_i  E_f } \int_L d^3 \bm{x}  \, e^{- i  \bm{q} \cdot \bm{x}} \,  \langle P_f, L \vert \mathcal O_B(0)   \vert \bar n, L \rangle \langle \bar n, L \vert      \mathcal J_A(0, \bm{x}) \vert P_i, L   \rangle  \,,
\label{eq:cbar}
\end{align}
one obtains
\begin{eqnarray}
\label{eq:Gdecom}
G_L(\tau, \bm{q}) \equiv  \sum_{n=0}^\infty     c_n \, \Theta(\tau) \,   e^{- [ E_n(L, \bm{P}_f + \bm{q}) - E_f ] \vert \tau \vert}   +  \sum_{\bar n =0}^\infty    \bar c_{\bar n}   \, \Theta(- \tau) \, e^{- [ E_{\bar n}(L, \bm{P}_i - \bm{q}) - E_i ] \vert \tau \vert} \,.
\end{eqnarray}
The finite-volume states $|n,L\rangle$ and $|\bar{n},L\rangle$ in Eqs.~(\ref{eq:c}) and (\ref{eq:cbar}), and therefore the $c_n $ and $\bar c_{\bar n} $ coefficients in Eq.~(\ref{eq:Gdecom}), differ in general not only due to differing three-momenta but also because the internal quantum numbers of the currents may be distinct. For example, processes like $K \to \pi \gamma$ are described by setting one current to the electromagnetic current and the other to the weak Hamiltonian, symbolically $\mathcal J_A = j_\mu$ and  $\mathcal O_B = \mathcal H_{W}$. For such a process, $c_{n,\mu}$ receives contributions from states with zero strangeness (such as $\pi \pi$ states) whereas $\bar c_{\bar n, \mu}$ contains intermediate states with strangeness equal to $-1$ (such as $K \pi$ states).

Depending on the detailed choices of states, currents and kinematics in Eq.~(\ref{eq:Gdecom}), finite-volume energies may exist for which $E_{\bar n}(L, \bm{P}_i - \bm{q})  \le  E_i - \omega $ or $E_n(L, \bm{P}_f + \bm{q}) \le E_f + \omega$, where $\omega$ is the energy carried away by the current $\mathcal{J}_A$. As a consequence, intermediate states can go on shell, generating the long-distance parts of these matrix elements. Such states are responsible for the dominant difference between finite-volume Euclidean and infinite-volume Minkowski correlation functions and are the focus of this work. To separate on- and off-shell states, it is useful to introduce cut-off indices, $N(\omega)$ and $\bar N(\omega)$, such that for $n  \geq N(\omega)$ and $\bar n  \geq \bar N(\omega)$, one has $ E_{\bar n}(L, \bm{P}_i - \bm{q}) + \omega > E_i $ and $ E_n(L, \bm{P}_f + \bm{q}) - \omega > E_f$ respectively, i.e.~the finite-volume intermediate states are off shell up to a current energy of $\omega$. Taking $N(\omega)$ and $\bar N(\omega)$ larger than the minimum requirements poses no problem, and as explained in more detail below, will likely be advantageous in practical implementations. 

With this discussion, one can define
 \begin{align}
 G_L^{< N}(\tau, \bm{q}) & \equiv   \sum_{n=0}^{N-1}     c_n \, \Theta(\tau) \,   e^{- [ E_n(L, \bm{P}_f + \bm{q}) - E_f ] \vert \tau \vert}   +  \sum_{\bar n =0}^{\bar N-1}    \bar c_{\bar n}   \, \Theta(- \tau) \, e^{- [ E_{\bar n}(L, \bm{P}_i - \bm{q}) - E_i ] \vert \tau \vert} \,, \\[3pt]
 \label{eq:TgN}
T_L^{\geq N}(\omega, \bm{q}) 
& \equiv   \int_{-\infty}^\infty d \tau \, e^{ \omega \tau}      \left[ G_L(\tau, \bm{q})-G_L^{< N}(\tau, \bm{q}) \right]   \,,
\end{align}
where to keep the notation simple, the superscripts $<N$ and $\geq N$ are taken to be the representative of both the $N$ and $\bar{N}$ dependence of the functions, and the $\omega$ dependence of $N$ and $\bar N$ is left implicit.  As mentioned in the introduction, the $c_n$  ($\bar{c}_{\bar{n}}$) coefficients for the intermediate states $0$ to $N-1$ ($\bar N - 1$) can be separately evaluated in a dedicated lattice-QCD calculation of three-point functions formed with optimized operators, see Eqs.~(\ref{eq:c}) and (\ref{eq:cbar}). In Eq.~\eqref{eq:TgN}, the subtracted integral is convergent by construction, as $N(\omega)$ and $\bar N(\omega)$ are chosen such that  $e^{ \omega \tau} \left[ G_L(\tau, \bm{q})-G_L^{< N}(\tau, \bm{q}) \right]$ decays with increasing $\vert \tau \vert$.  The result of the integration carries no memory of the Euclidean signature and thus brings us closer to the stated goal of recovering $\mathcal T$. However, the approach is clearly incomplete, since the intermediate states labeled from $0$ to $N-1$ ($\bar N-1$) are not accounted. Additionally,  the finite-$L$ effects have yet to be addressed.

One can now define
\begin{equation}
 T_L^{< N}(\omega, \bm{q})  \equiv \sum_{n=0}^{N-1}  \frac{ c_n }{E_n(L, \bm{P}_f - \bm{q})-(E_f + \omega)} + \sum_{\bar n =0}^{\bar N-1}  \frac{\bar c_{\bar n} }{E_{\bar n}(L, \bm{P}_i + \bm{q}) -  (E_i -\omega) } \,,
 \label{eq:TlN}
\end{equation}
which can naively be constructed using exactly the coefficients and energies defining $G_L^{< N}(\tau, \bm{q})$. As explained below, in practice one must take a slightly different approach to properly treat finite-volume effects. Note that $T_L^{\geq N}(\omega, \bm{q})$, defined in Eq.~\eqref{eq:TgN}, satisfies the same decomposition as that given here for $ T_L^{< N}(\omega, \bm{q})$, but with the sums running from $N, \bar N$ to $\infty$. Thus, the combination
\begin{equation}
T_L(\omega, \bm{q})  \equiv T_L^{< N}(\omega, \bm{q}) + T_L^{\geq N}(\omega, \bm{q}) \,, 
\label{eq:TLdef}
\end{equation}
gives the sum over all finite-volume poles from $0$ to $\infty$ and differs from the target quantity, $\mathcal T(\omega, \bm{q})$, only by the finite-volume effects encoded in $\Delta T_L(\omega, \boldsymbol q)$.

As stressed in the beginning of this section, the two-particle poles within $T^{< N}_L$ must be exactly canceled by those in $\Delta T_L$. This is essential because {(i)} these poles are artifacts of the particular volume, $L$, and cannot be part of the physical quantity $\mathcal T$, and {(ii)} the formalism holds for values of $\omega$ arbitrarily close to, indeed exactly coinciding with, the poles in a given lattice volume. Note that the same is not true for the poles within $T_L^{\geq N}(\omega, \bm{q}) $, which are safely above the range of allowed $\omega$ values.  The requirement of exact cancellation of divergences in $T^{< N}_L$ and $\Delta    T_L$ means that these two quantities must be treated in a consistent way. To explain this properly, we first need to give some details on the construction of $\Delta T_L(\omega, \boldsymbol q)$.

Both the generalized L\"uscher and Lellouch-L\"uscher formalisms are by now well understood, and here only the key equations are provided for completeness. First, the L\"uscher formalism provides a relationship between the finite-volume spectrum in the two-particle regime and the infinite-volume amplitude. This can be written as~\cite{Briceno:2014oea}
\begin{eqnarray}
\det\left[   F^{[x]}(L,P_n)^{-1}+\mathcal{M}^{[x]}(P_n)\right]=0 \,,
\label{eq:QC1}
\end{eqnarray}
where the superscript $[x]$ is set equal to either $\mathcal{J}\ket{P_i}$ or $\mathcal{O}\ket{P_i}$ and serves as an indicator of the internal quantum numbers for the two-particle states. Here, $\mathcal M^{[x]}$ is the infinite-volume $2 \to 2$ scattering amplitude and $ F^{[x]}$ is a known kinematic function, given explicitly in Eq.~\eqref{eq:FJJdef} of Sec.~\ref{sec:FV}. 
 As explained in detail in that section, the matrix and determinant space used here is a Kronecker-product space of (orbital angular momentum)$\otimes$(spin)$\otimes$(flavor channels). Depending on the time ordering of the currents, the quantization condition is to be evaluated at either $P_n = E_i - \omega$ or $P_n = E_f + \omega$,  where $q^\mu \equiv (\omega, \boldsymbol q)$ is the four-momentum of the current $\mathcal{J}_A$. One must then identify the set of solutions, in $P_n^0 = E_i - \omega$ or else $P_n^0 = E_f + \omega$, for which the left-hand side vanishes. This gives the finite-volume spectra, denoted by $ E_{\bar n}(L, \bm{P}_i - \bm{q})  $ (for $[x] = [\mathcal{J}\ket{P_i}] $) and $ E_n(L, \bm{P}_f + \bm{q})$ (for $[x]=[\mathcal{O}\ket{P_i}]$).

From the same building blocks that define the quantization condition, one can construct a matrix, $\mathcal F$, that plays an important role throughout this work:
\begin{eqnarray}
&& \mathcal{F}^{[ x]} (L,P) \equiv \frac{1}{{F^{[x]}}(L,P)^{-1} + \mathcal M^{[  x]}(P)} \,,
\label{eq:calF}
\end{eqnarray}
where $[x]$ and $P\equiv (E,\bm{P})$ are taken as generic representatives of the two-particle quantum numbers and momenta. This matrix has poles, $E_n$, whenever Eq.~\eqref{eq:QC1} is satisfied, and the corresponding residues define the generalized Lellouch-L\"uscher matrices~\cite{Briceno:2014uqa, Briceno:2015csa}
\begin{eqnarray}
\mathcal{R}_n^{[x]}(L,\bm{P}  )
&\equiv&\lim_{E  \rightarrow E_n} \left [ (E-E_n)~
\mathcal{F}^{[x]}(L,P)  \right ].
\label{eq:ampt-FV-correl-funct}
\end{eqnarray}

These factors allow one to relate finite- and infinite-volume matrix elements. For the present set-up the relevant relations are
\begin{align}
\label{eq:LL_formula1}
2 E_i L^6 \langle   n, L \vert      \mathcal O_B(0 ) \vert P_i, L   \rangle^2
& =  \mathcal{H}_{2 \to 1}^{[\mathcal{O}]}({P}_f + {q})\cdot \mathcal{R}_{  n}^{[\mathcal{O}\ket{P_i}]}(L,\boldsymbol {P}_f + \boldsymbol{q})  \cdot \mathcal{H}_{1 \to 2}^{[\mathcal{O}]}({P}_f + {q}) \,, \\
\label{eq:LL_formula2}
2 E_f L^6 \langle   P_f, L \vert      \mathcal J_A(0 ) \vert   n, L   \rangle^2
& =  \mathcal{H}_{2 \to 1}^{[\mathcal{J}]}({P}_f + {q})\cdot \mathcal{R}_{  n}^{[\mathcal{O}\ket{P_i}]}(L,\boldsymbol {P}_f + \boldsymbol{q})  \cdot \mathcal{H}_{1 \to 2}^{[\mathcal{J}]}({P}_f + {q}) \,, \\[5pt]
\label{eq:LL_formula3}
2 E_i L^6 \langle \bar n, L \vert      \mathcal J_A(0 ) \vert P_i, L   \rangle^2
& = \mathcal{H}_{2 \to 1}^{[\mathcal{J}]}({P}_i - {q}) \cdot \mathcal{R}_{\bar n}^{[\mathcal{J}\ket{P_i}]}(L,\boldsymbol {P}_i - \boldsymbol{q})  \cdot \mathcal{H}_{1 \to 2}^{[\mathcal{J}]}({P}_i - {q}) \,, \\
\label{eq:LL_formula4}
2 E_f L^6 \langle P_f, L \vert      \mathcal O_B(0 ) \vert \bar n, L   \rangle^2
& = \mathcal{H}_{2 \to 1}^{[\mathcal{O}]}({P}_i - {q}) \cdot \mathcal{R}_{\bar n}^{[\mathcal{J}\ket{P_i}]}(L,\boldsymbol {P}_i - \boldsymbol{q})  \cdot \mathcal{H}_{1 \to 2}^{[\mathcal{O}]}({P}_i - {q}) \,,
\end{align}
where various infinite-volume $1 + \mathcal J \to 2$ transition amplitudes are introduced, e.g., 
\begin{align}
\label{eq:MEdef1}
\mathcal H_{1 \to 2}^{[\mathcal O]} (P_f + q) &\equiv \langle P_f + q,\text{out} \vert \mathcal O_B(0) \vert P_i \rangle \,, 
\\
\label{eq:MEdef2}
\mathcal H_{2 \to 1}^{[\mathcal J]} (P_f + q) &\equiv \langle P_f  \vert \mathcal J_A(0) \vert  P_f + q,\text{in}\rangle \,,
\\[5pt]
\label{eq:MEdef3}
\mathcal H_{1 \to 2}^{[\mathcal J]} (P_i - q) &\equiv \langle P_i - q,\text{out} \vert \mathcal J_A(0) \vert P_i \rangle \,, 
\\
\label{eq:MEdef4}
\mathcal H_{2 \to 1}^{[\mathcal O]} (P_i - q) &\equiv \langle P_f  \vert \mathcal O_B(0) \vert  P_i - q,\text{in}\rangle \,.
\end{align}
The amplitudes defined with an ``out'' state are understood as column vectors and those with an ``in'' state as row vectors, acting on the space of angular momentum, spin and flavor of the two-hadron state.

\begin{figure}[t!]
\begin{center}
\includegraphics[scale=0.625]{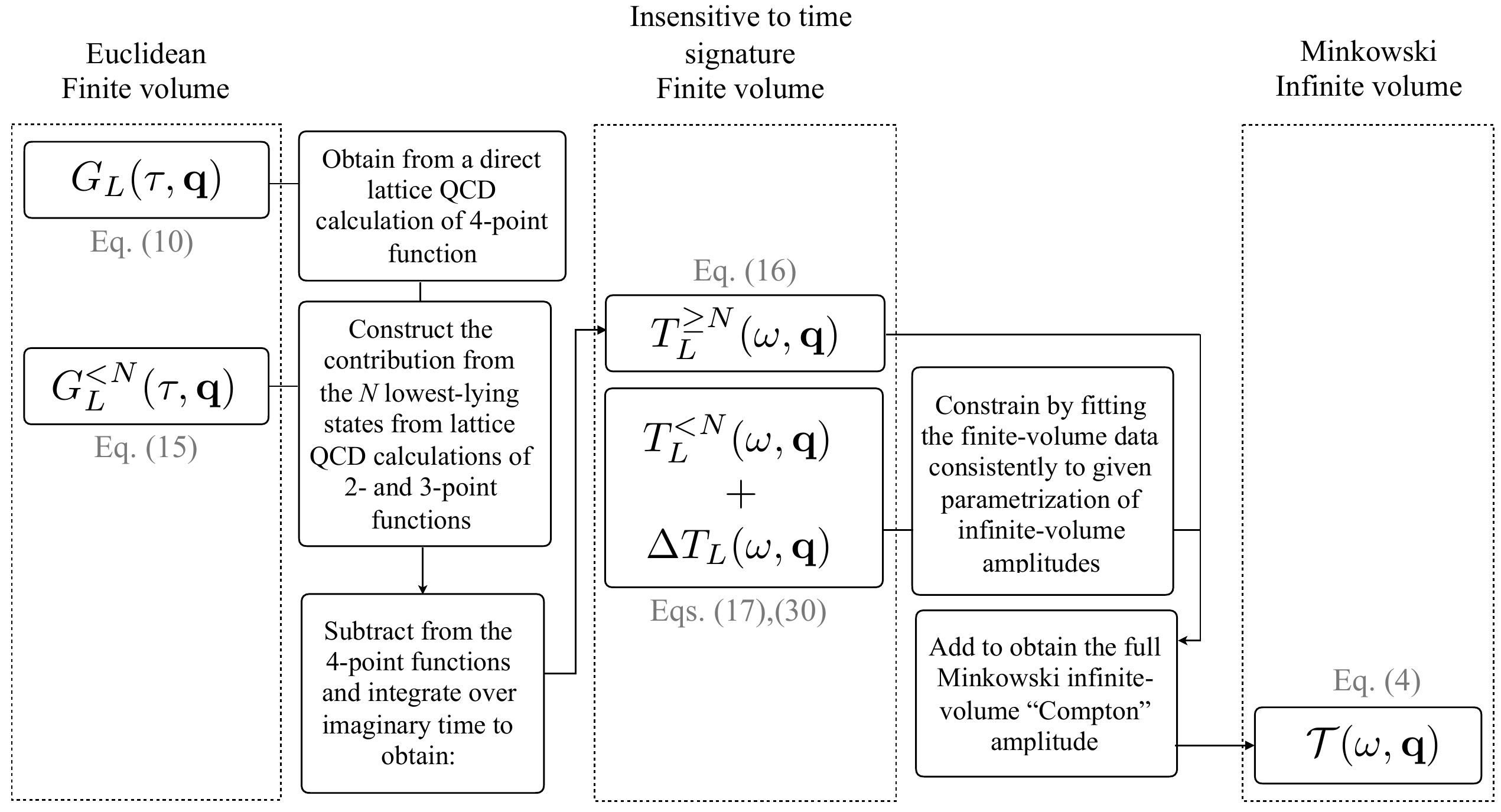}
\caption{$G_L(\tau, \bm{q})$ defined in Eq.~(\ref{eq:GFV}) is the starting point of matching the long-range bi-local matrix elements from lattice QCD to the infinite-volume Minkowski amplitude $\mathcal T(\omega, \bm{q})$. The intermediate quantities that must be obtained to fulfill this mapping, and the relations among them, are depicted in the chart. The $P_i$ and $P_f$ dependence of all functions is left implicit.
}
\label{fig:eqsflowchart}
\end{center}
\end{figure}
 
Now given a set of finite-volume energies, $E_{\bar n}(L, \bm{P}_i - \bm{q})$ and $E_n(L, \bm{P}_f + \bm{q})$, and the various finite-volume matrix elements noted above, one can constrain all quantities needed to remove the finite-volume effects from $T_L(\omega, \boldsymbol q)$, to extract $\mathcal T(\omega, \boldsymbol q)$. The correction is given by
\begin{align}
\begin{split}
\label{eq:fvcorrection}
 \Delta    T_L(\omega, \bm{q})    \ \equiv  \ & \mathcal{H}_{2 \to 1}^{[\mathcal J]}(P_f+q)   \cdot  \mathcal F^{[  \mathcal O \vert P_i \rangle]} (L,P_f+q)   \cdot  \mathcal{H}_{1 \to 2}^{[\mathcal O]}(P_f+q) \ +\
 \\[3pt]
 & \mathcal{H}_{2 \to 1}^{[\mathcal O]}(P_i-q)     \cdot  \mathcal F^{[  \mathcal J \vert P_i \rangle]} (L,P_i - q)  \cdot   \mathcal{H}_{1 \to 2}^{[\mathcal J]}(P_i-q)    \,,
 \end{split}
\end{align}
where each term on the right-hand side has the structure [row~vector]$\times$[matrix]$\times$[column vector], defined in the space of two-particle degrees of freedom.

In practice, the determination of the scattering amplitudes, $\mathcal M$, as well as the transition amplitudes, $\mathcal H$, will generally require fits of all finite-volume data to a set of parameterizations. Given a particular fit, one can then recover the finite-volume matrix elements, and thus the coefficients $c_n$ and $\bar c_{\bar n}$, as well as the corresponding finite-volume energies. It follows that the sum over low-lying poles, $T^{< N}_L(\omega, \boldsymbol q)$, can be evaluated in two ways, either directly from the finite-volume data, or from the energies and matrix elements predicted through a global fit. Given the statistical nature of the lattice-QCD data, for this analysis it is crucial that the second approach is used. Only in this way does one assure an exact cancellation of the poles in $\Delta T_L$ and $T^{<N}_L$ is reached. This requirement of a common parametrization for these two pieces is emphasized in Eq.~\eqref{eq:master} by the $\mathcal M,\mathcal H$ subscript. The cancellation of poles will also be illustrated in Sec.~\ref{sec:example} for a particular example. 

This completes the discussion of the procedure involved in extracting the infinite-volume amplitude $\mathcal T$ using the inputs that are defined in a finite Euclidean spacetime, as summarized in Eq.~\eqref{eq:master}. The quantities that must be evaluated in implementing this procedure, as well the relationships among them, are depicted in Fig.~\ref{fig:eqsflowchart} for further clarity.

\section{Finite-volume corrections to the four-point correlation function \label{sec:FV}}
\noindent
In this section, the expression for the additive finite-volume function, $\Delta    T_L(\omega, \bm{q})$, given in Eq.~\eqref{eq:fvcorrection}, is derived. The derivation requires a diagrammatic representation of finite-volume correlation functions, first laid out for systems with identical scalar particles by L\"uscher in Ref.~\cite{Luscher:1991n1}, and Lellouch and L\"uscher in Ref.~\cite{Lellouch:2000}, and presented in a purely quantum-field-theoretic context by Kim, Sachrajda, and Sharpe (KSS) in Ref.~\cite{Kim:2005gf}. The approach of KSS is also the starting point of Ref.~\cite{Christ:2015pwa} in the analysis of long-range effects in $K$-$\overline{K}$ mixing. These techniques have since been generalized to accommodate any number of open two-particle channels with arbitrary masses and spin~\cite{He:2005ey, Davoudi:2011md,Meyer:2011um,Hansen:2012tf,Briceno:2012yi, Briceno:2013lba, Briceno:2014oea,Briceno:2014uqa,Feng:2014gba, Briceno:2015csa}. Here we adopt this formalism to extend the work of Ref.~\cite{Christ:2015pwa}.

Consider the infinite-volume Minkowski-signature correlation function
\begin{align}
{\mathcal C}(P_f,P_i,q) & \equiv \int  d^4 x d^4 y d^4 z \, e^{ -  i P_i \cdot y +i q \cdot x  +  i P_f \cdot z} \langle 0 \vert  \,  \text{T} \{  {\Psi}'(z) i\mathcal J_A(x)  i\mathcal O_B(0) \Psi(y) \} \, \vert 0   \rangle \,,
\label{eq:calCdef}
\end{align}
where one of the four fields is left at the origin of position space to avoid the overall momentum-conserving delta function. The LSZ reduction formula implies that this correlation function has poles associated with the incoming and outgoing single-particle states, with quantum numbers set by $\Psi$ and $\Psi'$. By amputating the external legs and placing the momenta $P_i$ and $P_f$ on shell, one arrives at an alternative expression for $i\mathcal T$, first defined in Eq.~\eqref{eq:TAdef},
\begin{align}
i\mathcal{T}( \omega, \boldsymbol q)  
&= -  \lim_{P_i^0 \rightarrow E_i(\bm{P}_i),P_f^0 \rightarrow E_f(\bm{P}_f)} 
 \big ( P_i^2-M_i^2 \big )
 \big ( P_f^2-M_f^2 \big ) \,
{\mathcal C}(P_f,P_i,q) \,,
\label{eq:Tinfcorrdef}
\end{align}
where the fields are normalized such that $\langle P_i | \Psi(0) | 0 \rangle = \langle 0 |  {\Psi}'(0) | P_f \rangle = 1$, and as before the $P_i$ and $P_f$ dependence of the amplitude here and in Eq.~(\ref{eq:GL}) is left implicit.

The finite-volume counterpart of the correlation function in Eq.~(\ref{eq:calCdef}) can be written as
\begin{align}
C_L(P_f,P_i,q) & \equiv \int_L  d^4 x d^4 y d^4 z \, e^{ -  i P_i \cdot y +i q \cdot x  + i P_f \cdot z} \langle 0 \vert  \,  \text{T} \{  {\Psi}'(z) i\mathcal J_A(x)  i\mathcal O_B(0) \Psi(y) \} \, \vert 0   \rangle_L \,,
\label{eq:GL}
\end{align}
with the momentum and position coordinates still carrying Minkwoski signature.
The integrals over time coordinates are as above, but those over space coordinates are restricted to the finite cubic volume, as indicated by the $L$ subscript. A similar LSZ reduction can be implemented here to reach
\begin{align}
iT_L(\omega, \boldsymbol q) 
&\equiv -  \lim_{P_i^0 \rightarrow E_i(\bm{P}_i),P_f^0 \rightarrow E_f(\bm{P}_f)} 
 \big ( P_i^2-M_i^2 \big )
 \big ( P_f^2-M_f^2 \big ) \,
 C_L(P_f,P_i,q) 
\label{eq:GL} \,,
\\
\label{eq:TL_v2}
&=  2L^3\sqrt{E_iE_f}\, \int_L d^4 x \, e^{  i\omega t  -  i\bm{q} \cdot \bm{x}} \, \langle P_f,L \vert  \,  \text{T} \{ i\mathcal J_A(x)  i\mathcal O_B(0) \} \, \vert P_i,L   \rangle_{L,\text{conn.}}   \,.
\end{align}
This provides an alternative definition for $T_L(\omega, \boldsymbol q)$, the finite-volume counterpart of the Compton amplitude introduced in Eq.~\eqref{eq:TLdef}.

\begin{figure}[t]
\begin{center}
\includegraphics[width=0.95\textwidth]{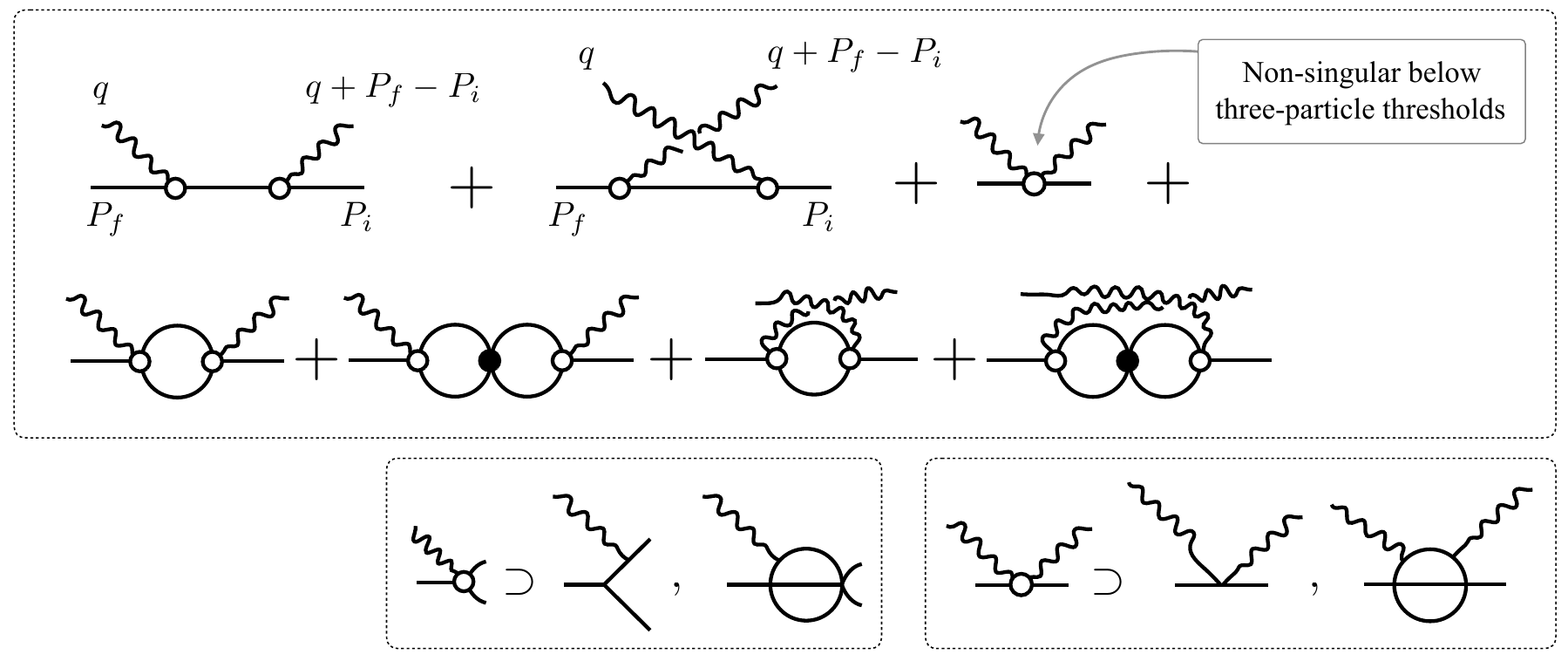}
\caption{A diagrammatic expansion of the Compton amplitude defined in Eqs.~\eqref{eq:calCdef} and \eqref{eq:Tinfcorrdef}. Here, the aim is to display explicitly all two-hadron intermediate states in the s-channel, as represented in Eq.~\eqref{eq:Tinfdecom}. In addition, single-particle intermediate states are separated, though these play a less important role in the finite-volume analysis of this section.
The two-current vertex, as well as the $1\to2$ kernel, contain no two-particle singularities below three-particle production thresholds. Representative contributions to the former and the latter are shown for a given theory in the bottom-right and bottom-left panels, respectively. 
The black circle represents the $2 \to 2$ scattering amplitude, $\mathcal M$.
}
\label{fig:comptonamp}
\end{center}
\end{figure}

At this point, the goal is to quantify the difference between Eqs.~\eqref{eq:Tinfcorrdef} and \eqref{eq:TL_v2}. These quantities correspond directly to the correlation functions studied by KSS~\cite{Kim:2005gf} and generalized by two of us in Ref.~\cite{Briceno:2015csa}. For completeness of the presentation, some details of the derivation will be repeated here. The first step is to derive a diagrammatic representation of the infinite-volume correlation function that explicitly displays all two-particle intermediate states. The result, illustrated in Fig.~\ref{fig:comptonamp}, can be represented algebraically in the following compact form: 
\begin{equation}
\label{eq:Tinfdecom}
 \mathcal T = \overline{\mathcal T} -\left [ \sum_{n=0}^\infty  \overline {\mathcal H}_{2 \to 1}^{[\mathcal J]} \left [ - \otimes \mathcal I \otimes \overline{\mathcal M}  \right ]^n    \otimes \mathcal I \otimes  \overline {\mathcal H}_{1 \to 2}^{[\mathcal O]}  + (\mathcal J \leftrightarrow \mathcal O)\right ]   \,,
\end{equation}
where we have chosen to work directly with $\mathcal T$ instead of $\mathcal C$ and have thus amputated all single-particle operator-dependent terms on both sides of the equations.
Here, five new building blocks are introduced: $ \overline{\mathcal T}$, $ \overline {\mathcal H}_{2 \to 1}^{[X]}$ , $ \otimes \mathcal I \otimes $, $ \overline{\mathcal M} $, and $  \overline {\mathcal H}_{1 \to 2}^{[X]}$ with $X = \mathcal J$ or $\mathcal O$, to be explained in turn. First, as a general rule, we use the overline to indicate a quantity in which all $s$-channel, two-particle-reducible diagrams have been discarded. Thus  $ \overline {\mathcal H}_{2 \to 1}^{[X]}$ and $  \overline {\mathcal H}_{1 \to 2}^{[X]}$ are generated by first developing a diagrammatic  expansion for Eqs.~\eqref{eq:MEdef1}-\eqref{eq:MEdef4}, and then discarding all diagrams that fall into disconnected pieces when any two lines, carrying the total four-momentum of the system, are cut. The quantity $\overline{\mathcal M}$, referred to as the Bethe-Salpeter kernel, is defined in exactly the same way, but with the $2 \to 2$ hadronic amplitude in place of $\mathcal H^{[X]}_{2 \to 1(1 \to 2)}$.\footnote{This quantity was named ${\mathcal B}$ in Ref.~\cite{Hansen:2019nir}.} The quantum numbers associated with $\overline{\mathcal M}$ and $\mathcal I$ (as well as those of $\mathcal S$ introduced below) are deduced from those of $ \overline {\mathcal H}_{2 \to 1 (1 \to 2)}^{[X]}$ present in each term.

In Eq.~\eqref{eq:Tinfdecom}, the various two-particle irreducible (2PI) quantities are combined via integrals over two-particle loops, denoted by $\otimes \mathcal I \otimes$. This symbol is thus built from two fully-dressed propagators, the requisite symmetry factor and the loop integral. The $\otimes$ symbol is meant to stress that the quantities are connected in a complicated way, in particular by integration over the loop momentum. This results in $ \overline {\mathcal H}_{2 \to 1}$, $ \overline{\mathcal M}$ and $  \overline {\mathcal H}_{1 \to 2}$ being evaluated at off-shell values of the momenta as well. As an explicit example, one has
\begin{multline}
\label{eq:calIdef}
\overline {\mathcal H}_{2 \to 1}^{[\mathcal J]}     \otimes i \mathcal I \otimes  \overline{\mathcal M}  \equiv \sum_a \xi_a  \int \frac{d^4 k}{(2 \pi)^4} \ \overline {\mathcal H}_{2 \to 1}^{[\mathcal J]} (P_f, q  \, ; \, k, P_f + q - k) \\
 \times \Delta_{a1}(k) \Delta_{a2}(P_f + q - k) \times  \overline{\mathcal M}     ( k, P_f + q - k \, ; \,  P_f ,q) \,,   
\end{multline}
where $\Delta_{a1}$ and $\Delta_{a2}$ are the fully dressed propagators for particles 1 and 2 in channel $a$, in accord with the quantum numbers of state $\mathcal O \vert P_i \rangle$. $\xi_a = 1/2$ if the particles in channel $a$ are identical and $1$ otherwise.  
Momentum dependences on the left and right of the semicolon in the arguments refer to the momenta of incoming and outgoing pairs of hadrons, respectively.

From the definitions of $\otimes \mathcal I \otimes$ and the various 2PI quantities, three important identities directly emerge:
\begin{align}
\label{eq:HandHbar-1}
&\mathcal H_{2 \to 1}^{[\mathcal J]}  = \overline {\mathcal H}_{2 \to 1}^{[\mathcal J]}    \sum_{n=0}^\infty   \left [ - \otimes \mathcal I \otimes \overline{\mathcal M}  \right ]^n \,, \\
\label{eq:HandHbar-2}
&\mathcal H_{1 \to 2}^{[\mathcal O]} =  \sum_{n=0}^\infty   \left [ - \overline{\mathcal M}  \otimes \mathcal I \otimes \right ]^n     \overline {\mathcal H}_{1 \to 2}^{[\mathcal O]} \,, \\
&   \mathcal M= \sum_{n=0}^\infty   \left [ - \overline{\mathcal M}  \otimes \mathcal I \otimes \right ]^n   \overline{\mathcal M}  =  \overline{\mathcal M}   \sum_{n=0}^\infty   \left [ - \otimes \mathcal I \otimes \overline{\mathcal M}  \right ]^n \,,
\end{align}
with analogous relations for $\mathcal{J} \leftrightarrow \mathcal{O}$ in Eq.~\eqref{eq:HandHbar-1}-\eqref{eq:HandHbar-2}. These are used in the derivation to express the finite-volume function, $T_L$, in terms of the physical amplitudes appearing on the left-hand sides.
Finally, $\overline {\mathcal T}$ in Eq.~(\ref{eq:Tinfdecom}) is defined by all diagrams not included in the other terms. This includes all connected 2PI contributions to $\mathcal T$ as well as diagrams that contain a single-particle intermediate state, as shown in Fig.~\ref{fig:comptonamp}. 

The purpose of the skeleton expansion for $\mathcal T$ is twofold. First, it allows to identify all sources of singularities, as well as imaginary contributions to $\mathcal T$. This will play a key role in Sec.~\ref{sec:check} where the unitarity of the amplitudes is established. Second, it enables identifying all power-like $L$-dependences in its finite-volume analog, $T_L$, defined in Eq.~\eqref{eq:GL}. In particular, as discussed in more detail in Refs.~\cite{Kim:2005gf,Hansen:2012tf,Briceno:2012yi, Briceno:2013lba, Briceno:2014oea,Briceno:2014uqa, Briceno:2015csa}, one can show that
\begin{equation}
\label{eq:TLdecom}
   T_L = \overline{\mathcal T} - \left [\sum_{n=0}^\infty  \overline {\mathcal H}_{2 \to 1}^{[\mathcal J]} \left [ - \otimes \mathcal S \otimes \overline{\mathcal M}  \right ]^n    \otimes \mathcal S \otimes  \overline {\mathcal H}_{1 \to 2}^{[\mathcal O]}  + (\mathcal J \leftrightarrow \mathcal O)\right]   \,.
\end{equation}
Here all quantities are as above except for $\otimes \mathcal S \otimes$. This is defined exactly as $\otimes \mathcal I \otimes$ but with the replacement $\int d^3 \boldsymbol k/(2 \pi)^3 \longrightarrow (1/L^3) \sum_{\boldsymbol k}$, see the example in Eq.~\eqref{eq:calIdef}. The sum runs over all three-momenta allowed by the finite-volume boundary conditions, in particular $\boldsymbol k = 2 \pi \boldsymbol n/L$ where $\boldsymbol n$ is a vector of integers.

Having set up the diagrammatic expansions for $T_L$ and $\mathcal T$, one can now substitute $\otimes \mathcal S \otimes = \otimes \mathcal I\otimes + F$, where $F$ is a finite-volume cut, defined by this relation and given more explicitly in Eq.~\eqref{eq:FJJdef} below. In words, the substitution rule reads: replace each sum with an integral plus a sum-integral difference, encoded by the symbol $F$. Two key simplifications then arise. First, the off-shell contributions from $\overline {\mathcal H}_{1 \to 2}$, $\overline{\mathcal M}$ and $\overline {\mathcal H}_{2 \to 1}$ lead to only exponentially suppressed volume effects that can be neglected. As a result, all functions adjacent to $F$ are projected to their on-shell momenta. Second, the propagators $\Delta_{a1}$ and $\Delta_{a2}$ are expanded about their on-shell point meaning that only the physical mass enters $F$. In particular, for spinning particles, the spin structure of the propagator becomes simple helicity projectors acting on the neighboring functions. 

\begin{figure}[t!]
\begin{center}
\includegraphics[width=.9\textwidth]{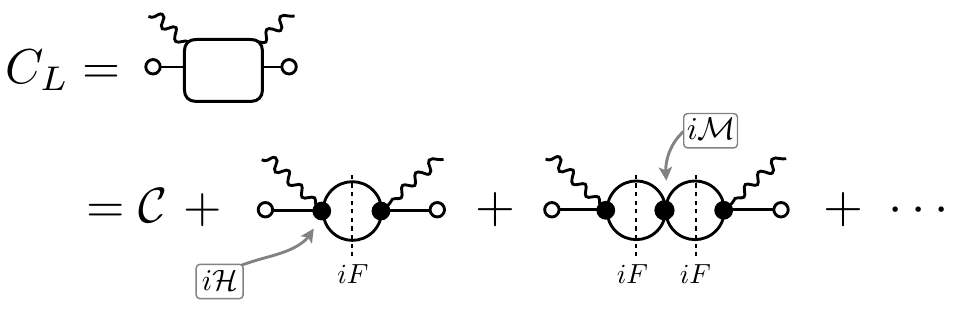}
\caption{ Shown is the full four-point correlation function in a finite volume. In the second line only the $s$-channel diagrams are shown explicitly, and the $u$-channel counterparts follow by switching the current vertices. The notation is similar to that used in Fig.~\ref{fig:comptonamp}. The vertical dashed lines denote contribution from the finite-volume $F$ function, defined in Eq~\eqref{eq:FJJdef}. The open circles denote the overlap of single-hadron state with the vacuum.}
\label{fig:CL}
\end{center}
\end{figure}

Substituting $\otimes \mathcal S \otimes = \otimes \mathcal I \otimes + F$ in Eq.~\eqref{eq:TLdecom} and summing all $\otimes \mathcal I \otimes$ dependences, and upon restoring the superscripts on all functions, one deduces
\begin{equation}
T_L = \mathcal T - \left [\sum_{n=0}^\infty  \mathcal H_{2 \to 1}^{[\mathcal J]} \left [ - F^{ [ \mathcal O \vert P_i \rangle]}   \mathcal M^{ [ \mathcal O \vert P_i \rangle]}  \right ]^n    F^{ [ \mathcal O \vert P_i \rangle]} \, \mathcal H^{[\mathcal O]}_{1 \to 2} + (\mathcal J \leftrightarrow \mathcal O)\right]   \,. 
\end{equation}
This is the main result of the derivation, analogous to Eqs.~(44) and (84) of Refs.~\cite{Kim:2005gf} and \cite{Briceno:2015csa}, respectively. Summing the geometric series in $n$ then directly gives Eq.~\eqref{eq:fvcorrection} of the previous section.

At this point it remains only to define $F^{[X \vert P_i \rangle]}$, the finite-volume cut matrix defined on the Kronecker-product space of (orbital angular momentum)$\otimes$(spin)$\otimes$(flavor channels).\footnote{The Kronecker-product symbol should not be confused with the same symbol used in this section to define loop integrals or sums.}
In order to  provide an explicit form, one can change to the basis of total angular momentum and use the compact notation introduced in Ref.~\cite{Briceno:2015csa}. Let $Jm_J$ denote the total angular momentum and its azimuthal component and $S$ and $\ell$ the total spin and orbital angular momentum. Also, label the channel space using indices $a$ and $a'$.   Then the $F^{[X \vert P_i \rangle]}$ function can be written as~\cite{Briceno:2014oea,Briceno:2015csa}
\begin{align}
\label{eq:FJJdef}
F^{[X \vert P_i \rangle]}_{\{J\},\{J'\}}(P,L) & \equiv  
\xi_a \delta_{aa'} 
\delta_{SS'}  \sum_{m_\ell,m'_\ell,m_S} \langle \ell \,m_\ell, S \, m_S \vert J m_J \rangle  \langle \ell' \,m_\ell', S' \, m_S \vert J' m_J' \rangle  
\nn\\
&\hspace{1.5cm}
\times \bigg [ \frac{1}{L^3} \sum_{\boldsymbol k } - \int \frac{d\boldsymbol k}{(2 \pi)^3} \bigg ]
\, 
\frac{   4\pi \,{Y}_{\ell m_\ell}(\hat {\boldsymbol k}^*_a) \, {Y}^*_{\ell'm'_\ell}(\hat {\boldsymbol k}^*_a)  }{2 \omega_{a 1} 2 \omega_{a 2}(E -  \omega_{a 1} - \omega_{a 2} + i \epsilon )} \left(\frac{k_a^*}{q_a^*}\right)^{\ell +\ell'} \,,
\end{align}
where $\xi_a$ is the symmetry factor defined after Eq.~\eqref{eq:calIdef}. Note that the superscript $[X \vert P_i \rangle]$ on $F$ is related to the quantum numbers of allowed two-hadron channels in this relation. The particles in channel $a$ are labeled with the numbers $1$ and $2$ and their corresponding masses are $m_{a1}$ and $m_{a2}$. Then the magnitude of back-to-back three-momentum in the center-of-mass frame is given by $q_a^*$, which is the solution to
\begin{equation}
E^* = \sqrt{m^2_{a1} + q_a^{*2}} + \sqrt{m^2_{a2} + q_a^{*2}} \,.
\end{equation}
 In Eq.~\eqref{eq:FJJdef}, we have also introduced $\omega_{a 1}=\sqrt{\boldsymbol k^2+m_{a1}^2}$, $\omega_{a 2}=\sqrt{(\boldsymbol P - \boldsymbol k)^2+m_{a2}^2}$, and $\boldsymbol k^*_a = k^*_a \hat {\boldsymbol k}^*_a$, with the latter defined as the spatial part of the four-vector reached by boosting $(\omega_{a1}, \boldsymbol k)$ with boost velocity $\boldsymbol \beta = - \boldsymbol P/E$. $E$ is the total energy of two hadrons in the lab frame, i.e., $E=\gamma E^*$.\footnote{The $*$ superscript, when applied to an energy or momentum, indicates that the quantity is defined in the center-of-mass frame. By contrast, when applied to a spherical harmonic or other generic function, the $*$ indicates  complex conjugate.} Equation~\eqref{eq:FJJdef} provides the final missing piece for the main result of this work, Eq.~\eqref{eq:master}.

 The other necessary ingredients, namely the generalized L\"uscher  and  Lellouch-L\"uscher formalisms have been previously derived in Refs.~\cite{Luscher:1991n1, Rummukainen:1995vs, Kim:2005gf, He:2005ey, Davoudi:2011md, Hansen:2012tf, Briceno:2012yi, Briceno:2013lba, Briceno:2014oea,Lellouch:2000, Meyer:2011um, Briceno:2012yi, Briceno:2014uqa, Feng:2014gba, Briceno:2015csa, Briceno:2015tza, Baroni:2018iau}. For completeness, here we provide a re-derivation of the most general results, directly from Eq.~(\ref{eq:fvcorrection}). First, the generalized L\"uscher formalism follows from the observation that poles in $\Delta T_L$ correspond to the finite-volume energies of the theory. As already discussed in the text around Eq.~\eqref{eq:calF}, this occurs whenever $\mathcal F^{[x]}(L,P)$ has a divergent eigenvalue, leading directly to Eq.~\eqref{eq:QC1}.

 Second, to obtain the generalized Lellouch-L\"uscher relation, one must match the residues at the poles defining $  T_L$, determined in two different ways. To this end, consider a special case of Eq.~\eqref{eq:TL_v2} for which $\mathcal{J}=\mathcal{O}$, the initial and final states are the same and $P_i=P_f$.   In this case, one can first aim for an expression for $  T_L$ in terms of finite-volume matrix elements.  This follows directly from Eqs.~\eqref{eq:c}, \eqref{eq:cbar}, \eqref{eq:TlN} and \eqref{eq:TLdef}. Choosing the kinematics so that $E_i - \omega$ is arbitrarily close to a finite-volume energy, $E_n$, one obtains 
 \begin{align}
\lim_{E_i-\omega \rightarrow E_n}  T_L(P_i,q) 
&= 
- \frac{2 E_i L^6     \langle  P_i, L   \vert   \mathcal J(0) \vert \bm{P}_i - \bm{q},E_n \rangle \langle \bm{P}_i - \bm{q},E_n \vert   \mathcal J(0)    \vert P_i, L   \rangle}
{(E_i-\omega)-E_n} \,.
\end{align}
To reproduce a second expression for $T_L$ near the finite-volume pole, one notes that the poles are encoded in Eq.~\eqref{eq:fvcorrection} and thus
\begin{align}
\lim_{E_i-\omega \rightarrow E_n} T_L(P_i,q) 
& =  
- \lim_{E_i-\omega \rightarrow E_n}
\left[
\mathcal{H}_{1 \to 2}^{[\mathcal{J}]}(P_i-q)
\mathcal{F}^{[\mathcal{J}\ket{P_i}]}(L,P_i-q) 
\mathcal{H}_{2 \to 1}^{[\mathcal{J}]}(P_i-q)\right].
\label{eq:iTL_nearpole}
\end{align}
Equating these two expressions reproduces Eqs.~\eqref{eq:LL_formula1}-\eqref{eq:LL_formula4} (upon trivial assignments of the kinematic arguments and the operator types). Note that in Eqs.~\eqref{eq:LL_formula1}-\eqref{eq:LL_formula4} the matrix elements are squared. This follows from our assumption that the current is Hermitian up to a possible CP-violating phase, so that one can write $\mathcal J(0) = e^{- i \alpha} \mathcal J_{\text H}(0) $ with $ \mathcal J_{\text H}(0) =  \mathcal J_{\text H}(0)^\dagger$. If we then further adopt a phase convention on the states such that the Hermitian current has a real finite-volume matrix element, it follows that
\begin{multline}
  \langle  P_i, L   \vert   \mathcal J(0) \vert \bm{P}_i - \bm{q},E_n \rangle \langle \bm{P}_i - \bm{q},E_n \vert   \mathcal J(0)    \vert P_i, L   \rangle  = \\  e^{- 2 i \alpha} \langle \bm{P}_i - \bm{q},E_n \vert   \mathcal J_{\text H}(0)    \vert P_i, L   \rangle^2  
   =  \langle \bm{P}_i - \bm{q},E_n \vert   \mathcal J(0)    \vert P_i, L   \rangle^2  \,.
\end{multline}

Before closing this section, we elaborate further on the scope of the results derived. In particular, we note that the general angular momentum, spin and flavor structure of the finite-volume function, $F$, defined in Eq.~(\ref{eq:FJJdef}), is a new feature 
that substantially complicates 
the extraction of Compton-like amplitudes using lattice QCD. 
However, the main challenges already arise in the determination of coupled hadronic scattering amplitudes (via the generalized L\"uscher formalism) as well as one-to-two transition amplitudes (using the generalization of the Lellouch-L\"uscher relation). As a result of significant numerical and algorithmic progress in recent years, implementation of these ideas is already well underway.

The central complication, whenever multiple flavor or angular-momentum channels need to be considered, is that one no longer has a direct one-to-one mapping between finite- and infinite-volume observables. In such cases, the only known approach is to determine a large set of finite-volume energies and matrix elements, for multiple choices of box size and total momentum, and to subsequently perform a global amplitude analysis using the relevant finite-volume formalism. This was first implemented for hadronic amplitudes with partial waves unphysically mixing in the finite volume \cite{Dudek:2012gj}, and was later extended to cases where multiple flavor channels are kinematically allowed and must be disentangled \cite{Dudek:2014qha,Wilson:2015dqa, Briceno:2016mjc,Dudek:2016cru}. Most recently, the methods have been applied to particles with intrinsic spin, inducing also an infinite-volume coupling between different orbital angular momenta \cite{Woss:2018irj, Woss:2019hse} and to transition amplitudes in which states are coupled via the spin and momentum of the external current \cite{Briceno:2016kkp,Briceno:2015dca}.
This progress provides strong empirical evidence that there is no significant obstacle for implementing the proposed formalism for Compton-like amplitudes in the kinematic region considered in this work.

\section{Unitarity check\ ~\label{sec:check}}
\begin{figure}[t!]
\begin{center}
\includegraphics[width=.95\textwidth]{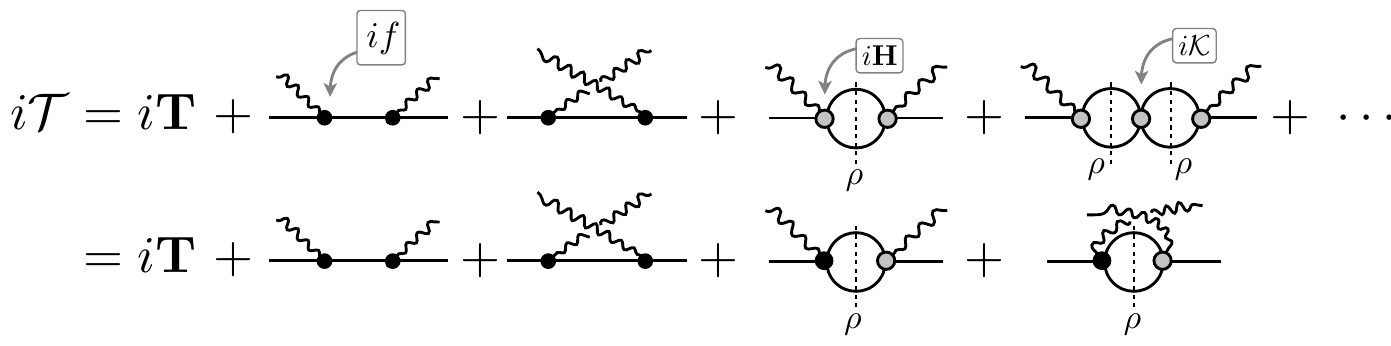}
\caption{ Shown is the representation of the Compton-like amplitude to all orders in perturbation theory, in which the contributions from imaginary parts (proportional to the phas-space factor $\rho$) are isolated. The black dots are either the full $1 \to 1$ vertex with a single current insertion, $if$, or the $1 \to 2$ or $2 \to 1$ scattering amplitudes, $i \mathcal H$. The gray circles define, $i\textbf H$ and $i \mathcal K$, respectively, which are the counterparts of transition and scattering amplitudes when the principal-valued prescription has been used in the s-channel loops. Further detail on these quantities is given the text.}
\label{fig:icalT}
\end{center}
\end{figure}

This section provides an additional check on Eq.~\eqref{eq:master} by proving that the result is consistent with the S-matrix unitarity.  As was recently shown in the context of three-particle scattering in Ref.~\cite{Briceno:2019muc}, all-orders perturbation theory can be used to directly extract all imaginary contributions to the scattering amplitude in a given kinematic region. Since the diagrammatic description emerges from a unitary theory, the result of summing over all contributions must automatically respect any consequence following from the unitarity relation, i.e., $S^\dagger S = 1$, both in the finite and infinite volume.  Nevertheless,  it is instructive to see how the final result directly satisfies the expected constraint. 

Let us first work out the consequences of unitary for $\mathcal T$. To this end, consider an all-orders expression for the Compton amplitude in the infinite volume, as depicted in Fig.~\ref{fig:icalT}. For the purpose of this check, the two local currents are assumed to be the same and to be Hermitian ($\mathcal J_A = \mathcal O_B$), and the initial and final states are set to coincide, with $P_i = P_f$.\footnote{As discussed above, our definition of $\mathcal T$ does not include the disconnected contribution that arises in this case.} When necessary, the dependence on the kinematic variable $\sqrt{s} = E^*$ is made explicit throughout this section, while the full dependence of the functions on kinematic variables, including on the momentum transfer of the currents, will be suppressed for brevity.
 In the kinematic region in which only two-particle states can go on shell, one can analyze all diagrams by systematically isolating their imaginary contributions. These are present only in the two-particle s-channel loops and are proportional to the phase-space factor $\rho$, defined as 
\begin{align}
\rho_a(s) & \equiv \frac{ q^{*}_a(s)\,\xi_a}{8\pi \sqrt{s}} \,.
\label{eq:ps}\end{align}

The phase-space factor, $\rho_a(s)$, arises from the imaginary part of the $i \epsilon$ pole prescription in the two-particle loops. To capture all contributions to the amplitude, it is convenient to split the full $i \epsilon$ integral into its real and imaginary parts. Performing this for a purely hadronic $2 \to 2$ scattering amplitude yields the standard relation 
\begin{align}
\label{eq:Kmat}
i\mathcal{M}(s) &= i\mathcal{K}(s) \frac{1}{1-i\rho(s) \mathcal{K}(s) } \,, \end{align}
where $\mathcal K$ is the K matrix, defined to have the same diagrammatic expansion as the scattering amplitude, but evaluated with the principal-value prescription for the s-channel loops. 
Similarly, the principal-valued version of $\mathcal{H}_{1 \to 2}$ can be denoted by $\mathbf{H}_{1 \to 2}$, which is related to $\mathcal{H}_{1 \to 2}$ via 
\begin{align}
\label{eq:Hmat}
i\mathcal{H}_{1 \to 2} &=   i\mathbf{H}_{1 \to 2} \frac{1}{1-i  \mathcal{K}(s)  \rho(s) } \,.
\end{align}
A similar relation for $\mathcal{H}_{2 \to 1}$ can be realized in terms of $\mathbf{H}_{2 \to 1}$,
\begin{align}
\label{eq:Hmat}
i\mathcal{H}_{2 \to 1} &=   \frac{1}{1-i  \mathcal{K}(s)  \rho(s) }  i\mathbf{H}_{2 \to 1} \,,
\end{align}

As with the various finite-volume quantities in the previous section, Eqs.~\eqref{eq:Kmat}-\eqref{eq:Hmat} are built from matrices ($\rho$, $\mathcal K$ and $\mathcal M$) and column vectors ($\mathcal{H}_{1 \to 2}$ and $\mathbf{H}_{1 \to 2}$) on the Kronecker-product space of (orbital angular momentum)$\otimes$(spin)$\otimes$(flavor channels).
In the case of two scalar particles in a single flavor channel, this space reduces to only angular momentum and all matrices become diagonal. Then one can draw a correspondence between a given entry of $\mathcal H_{1 \to 2}$ and $\mathcal M$. In particular, one finds
\begin{equation}
\frac{\text{Im} \, \mathcal M^{(\ell)}(s)}{\text{Re} \, \mathcal M^{(\ell)}(s)} = \frac{\text{Im} \,  \mathcal{H}_{1 \to 2}^{(\ell)}(s) }{\text{Re}\, \mathcal{H}_{1 \to 2}^{(\ell)} (s)} \,,
\end{equation}
meaning that the amplitudes have the same complex phase in each partial wave. This is Watson's theorem, and by returning to the general matrix space, one recovers a simple generalization for the case of multiple two-particle channels~\cite{Briceno:2014uqa}. 

In direct analogy to these more standard results, one can further determine an expression for the Compton amplitude that is consistent with unitarity to all orders in perturbation theory. This, in the kinematic region below three-hadron production thresholds, divides into a purely real principal-valued version of $\mathcal T$, denoted $\mathbf T$, as well as a series of $\rho$ cuts, analogous to those appearing in Eqs.~\eqref{eq:Kmat}-\eqref{eq:Hmat}, above. In addition, one must include a single-particle piece represented by the second and third terms on the right-hand side of Fig.~\ref{fig:icalT}. To express this, let us denote by $f(q^2)$ the $1 \to 1$ matrix element with a single current insertion and label the corresponding particle mass by $m$. Combining all terms gives
\begin{align}
\label{eq:compton}
i\mathcal{T} &= \left [i\mathbf{T} + 
if(q^2)   \, \frac{ \, i \, }{s - m^2+i\epsilon}
if(q^2)
+ i\mathbf{H}_{2 \to 1}
\frac{1}{1-i\rho(s) \mathcal{K}(s) }\rho(s) \,
i\mathbf{H}_{1 \to 2}\right ]
+[s \leftrightarrow u] \,,
\end{align}
where $s = (P-q)^2$ and $u = (P+q)^2$ are the usual Mandelstam variables, and the last term indicates the inclusion of terms in which the incoming and outgoing currents are interchanged, as shown in Fig.~\ref{fig:icalT}.

At this point, it is apparent that $i\cal T$ shares certain features with the previously considered amplitudes, $\mathcal M$ and $\mathcal H_{1 \to 2~(2 \to 1)}$. First, besides the single-particle pole, the only singularity in the amplitude arises from the phase-space factor $\rho_a(s) \sim \sqrt{s-(m_{a1}+m_{a2})^2}$. This threshold singularity is common to all three amplitudes. Second, in the case that a single two-particle channel is open, the third term in Eq.~\eqref{eq:compton} breaks into a sum over the angular momentum modes and the phase of each term matches that of $\mathcal{M}^{(\ell)}$ and $\mathcal{H}^{(\ell)}$.

Below the lowest-lying two-particle threshold, $i \rho$ and thus also $\mathcal M$ and $\mathcal H_{1 \to 2~(2 \to 1)}$ become purely real.
Similarly, the only possible imaginary contribution to the Compton amplitude below threshold is due to the $i \epsilon$ prescription within the single-particle pole. Above the lowest lying two-particle threshold, $i \rho$, $\mathcal{M}$, $\mathcal{H}_{1 \to 2(2 \to 1)}$, and $\mathcal{T}$ are all complex valued. In particular, from Eqs.~\eqref{eq:Kmat}-\eqref{eq:Hmat}, and \eqref{eq:compton} one can readily show that
\begin{align}
\label{eq:Munit}
& \text{Im} \, \mathcal{M}(s)   = \mathcal{M}^*(s) \, \overline{\rho}(s) \, \mathcal{M}(s) \,, \\[5pt]
& \text{Im} \, \mathcal{H}_{1 \to 2}     =     \mathcal H_{1 \to 2} \overline{\rho}(s) \, \mathcal{M}^*(s) \, \,,~ \text{Im} \, \mathcal{H}_{2 \to 1}    = \mathcal{M}^*(s) \, \overline{\rho}(s) \,    \mathcal H_{2 \to 1} \,, \\[5pt]
& \text{Im} \, \mathcal{T}    =
 \mathcal{H}^*_{2 \to 1} \, \overline{\rho}(s) \, \mathcal{H}_{1 \to 2}
 +(s \leftrightarrow u) \,,
\end{align}
where $\overline{\rho}_a(s) \equiv \rho_a(s)\, \Theta(s-(m_{a1}+m_{a2})^2)$, with $\Theta$ being the Heaviside step function.

Having established the unitarity constraint on the various infinite-volume amplitudes entering the formalism, it can now be shown that Eq.~\eqref{eq:master} is consistent with these expressions. To this end, it is sufficient to demonstrate ${\rm Im} \, \Delta T_L  = {\rm Im} \, \mathcal T$, which, together with the fact that $G^{\geq N}$ and $T^{< N}$ are real, demonstrates the desired consistency. One first notes that
\begin{align}
{\rm Im} \,\Delta T_L  
&=
{\rm Im}\left[\mathcal{H}_{2 \to 1}
\,
\,\mathcal{F}_L 
\,\mathcal{H}_{1 \to 2}\right]
+(s \leftrightarrow u) \,,
\\
&=
\left(\mathcal{H}_{2 \to 1} \mathcal{M}^{-1}
\right)^*
{\rm Im} \bigg [\mathcal{M}
F 
 \frac{1}{\mathcal{M}^{-1} + F} 
\bigg]
\left(\mathcal{M}^{-1} \mathcal{H}_{1 \to 2}\right)
+(s \leftrightarrow u) \,,
\label{eq:ImDT3}
\end{align}
where all $s$ dependences are dropped for brevity. Here, we use the definition $\text{Im} X \equiv (X - X^\dagger)/(2i)$ for a generic matrix, $X$, together with the fact that $\mathcal{H}_{2 \to 1} \mathcal{M}^{-1}$ and $\mathcal{M}^{-1} \mathcal{H}_{1 \to 2}$ are real-valued 
and equal, entry by entry.
The next step is to make use of the fact that $\mathcal{M}^{-1} + F$, and thus also its inverse, are Hermitian matrices. This leads to the substitution
\begin{align}
2 i \ {\rm Im} \bigg [\mathcal{M}
F 
 \frac{1}{\mathcal{M}^{-1} + F} \bigg ]  & = \mathcal{M}
F 
 \frac{1}{\mathcal{M}^{-1} + F} -    \frac{1}{\mathcal{M}^{-1 \dagger} + F^\dagger} F^\dagger \mathcal M^\dagger  \\
 & = \big [\mathcal M F - \mathcal M^\dagger F^\dagger \big ]  \frac{1}{\mathcal{M}^{-1} + F}  \,,
 \label{eq:Imp}
\end{align}
where, in the second line, we have used $[\mathcal{M}^{-1 \dagger} + F^\dagger]^{-1}  F^\dagger \mathcal M^\dagger = \mathcal M^\dagger  F^\dagger  [\mathcal{M}^{-1 \dagger} + F^\dagger]^{-1} $.
Next one observes that
\begin{align}
\frac{ \mathcal M F - \mathcal M^\dagger F^\dagger }{2i}& 
= \text{Im} [\mathcal M] F + \mathcal M^\dagger \text{Im} [ F ] \,  = \mathcal M^* \bar \rho \mathcal M F + \mathcal M^* \bar \rho = \mathcal M^* \bar \rho \mathcal M \, [\mathcal M^{-1} + F]  \,,
\label{eq:Imp2}
\end{align}
where $\mathcal M^\dagger = \mathcal M^*$, the unitary constraint on $\mathcal M$, Eq.~(\ref{eq:Munit}), as well as the relation ${\rm Im}[F]=-{\rm Im}[\mathcal M^{-1}]=\overline{\rho}$ are used. The use of Hermitian conjugation in the definition of $\text{Im}$ is mainly motivated by this final step, since due to the presence of spherical harmonics in the definition of $F$, $(F - F^*)/(2i) \neq \overline \rho$.  Finally, substituting Eqs.~\eqref{eq:Imp} and \eqref{eq:Imp2} into Eq.~\eqref{eq:ImDT3} leads to
\begin{align}
{\rm Im} \,\Delta T_L  &=
\mathcal{H}_{2 \to 1}^*\, 
\overline{\rho} \, \mathcal{H}_{1 \to 2}
+(s \leftrightarrow u)  = {\rm Im} \, \mathcal T \,,
\label{eq:imDeltaTL}
\end{align}
as desired.

In summary, we have provided an expression for the Compton amplitude, in terms of purely real building blocks, that is exactly consistent with unitarity, and have used this to express $\text{Im} \, \mathcal T$ in terms of transition amplitudes and phase-space factors. This expression is shown to be consistent with Eq.~\eqref{eq:master}, the main result of this work. The decomposition of $\mathcal T$ is built from the single-particle form factor and propagator, the phase-space factor $\rho$, the amplitudes $\mathcal H_{1 \to 2}$, $\mathcal H_{2 \to 1}$, and $\mathcal M$, as well as a real-valued short distance piece, denoted by $\mathbf T$. In fact, since all but the last function can be obtained from two- and three-point correlation functions in a finite-volume study, it is natural to think of $\mathbf T$ as the target observable, determination of which requires the method summarized by Eq.~\eqref{eq:master}.
In the next section we discuss this point in further detail, and highlight one subtlety that arises when considering amplitudes for which the corresponding K matrix has real-valued poles. 

\section{Numerical implementation~\label{sec:example}}
\noindent

 \begin{figure}[t]
\begin{center}
\includegraphics[width=.85\textwidth]{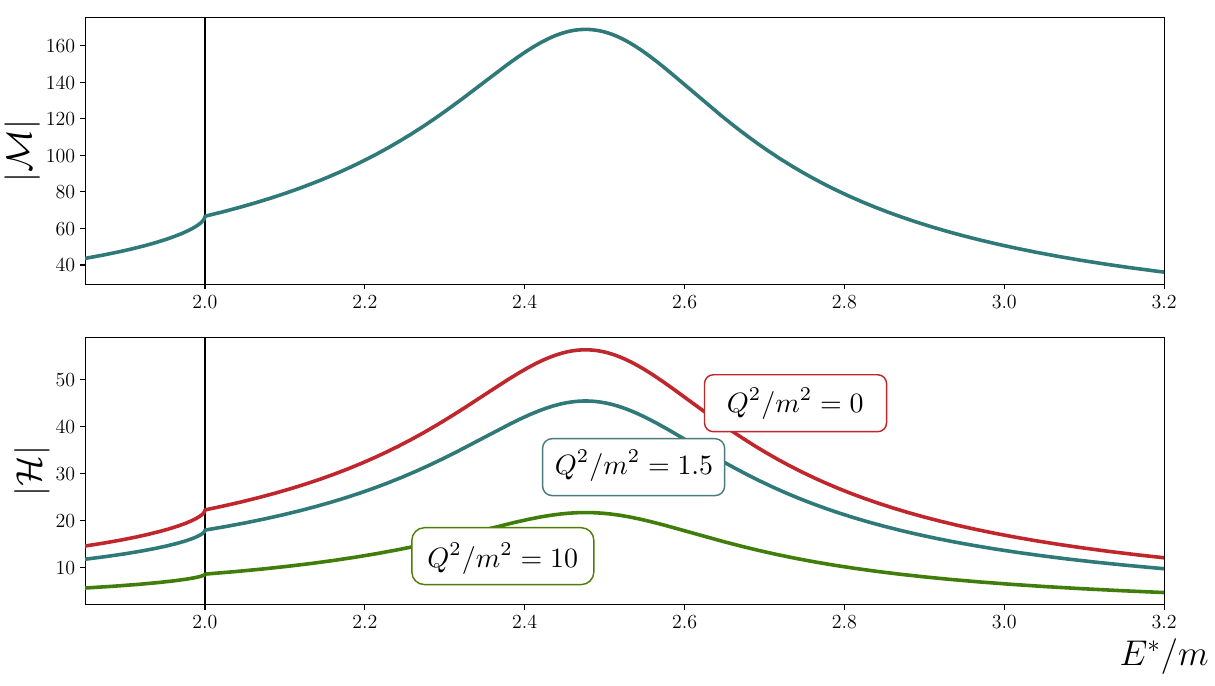}
\caption{Magnitude of the scattering amplitude (upper panel) and $1+\mathcal{J} \to 2$ transition amplitude (lower panel) using $^{m_R}/_m= 2.5$ and $g=3.0$ in the functions defined in the text. Three representative values of momentum transfer are shown for the transition amplitude. }
\label{fig:M_H}
\end{center}
\end{figure}

In this section, we provide a numerical example to illustrate the approach outlined in Sec.~\ref{sec:main_result} above and summarized by Eq.~\eqref{eq:master}. To reduce technical complications, the example involves kinematics for which only a single channel is open. The channel is composed of two identical scalars, each with mass $m$. It is further assumed that only the lowest partial wave ($\ell=0$) contributes. The initial and final single-hadron states are taken to be the same and are put at rest. Finally, the external current is taken to be a scalar, and as a result, the form factor $f(Q^2)$ and transition amplitude $\mathcal H(s, Q^2)$ are scalar functions. The subscripts $1 \to 2$ and $2 \to 1$ on $\mathcal{H}$ will be dropped in this section. The dependence on kinematic variables is displayed using a Lorentz-invariant notation, where $s=(P-q)^2$ is the invariant mass of the system and $q^2=-Q^2$, where $q$ is the momentum transferred by the first current.

As stressed in the previous section, the formalism can only be applied to amplitudes that exactly satisfy unitarity. In the case of the hadronic scattering amplitude, $\mathcal M(s)$, this means one must use Eq.~\eqref{eq:Kmat} with a real-valued $\mathcal K(s)$. With the standard relation to the scattering phase shift $\delta(q^*)$,
\begin{align}
{\cal{K}}(s)&=\frac{16\pi \sqrt{s}}{ q^{*}} \frac{1}{\cot\delta(q^{*})} \bigg \vert_{q^{*2} = s/4 - m^2} \,,
\end{align}
and the use of a Breit-Wigner parametrization for the latter,
\begin{align}
\tan \delta(q^*) &= \frac{ \sqrt{s} \, \Gamma(s)}{m_R^2 - s}  \,, \qquad \Gamma(s) = \frac{g^2}{6\pi} \frac{m_R^2}{s} q^* \,.
\end{align}
$\mathcal{K}$ is fully constrained. Here, $m_R$ is related to the mass of the resonance and $g$ characterizes the coupling to two-particle states. In the numerical results, the following values are assumed: ${m_R} = 2.5\,m$ and $g=3.0$. As can be seen in Fig.~\ref{fig:M_H}, these values lead to a standard peak-like structure near $s = m_R^2$ and a visible cusp at threshold, $s = 4 m^2$. 

  \begin{figure}[t]
\begin{center}
\includegraphics[width=1\textwidth]{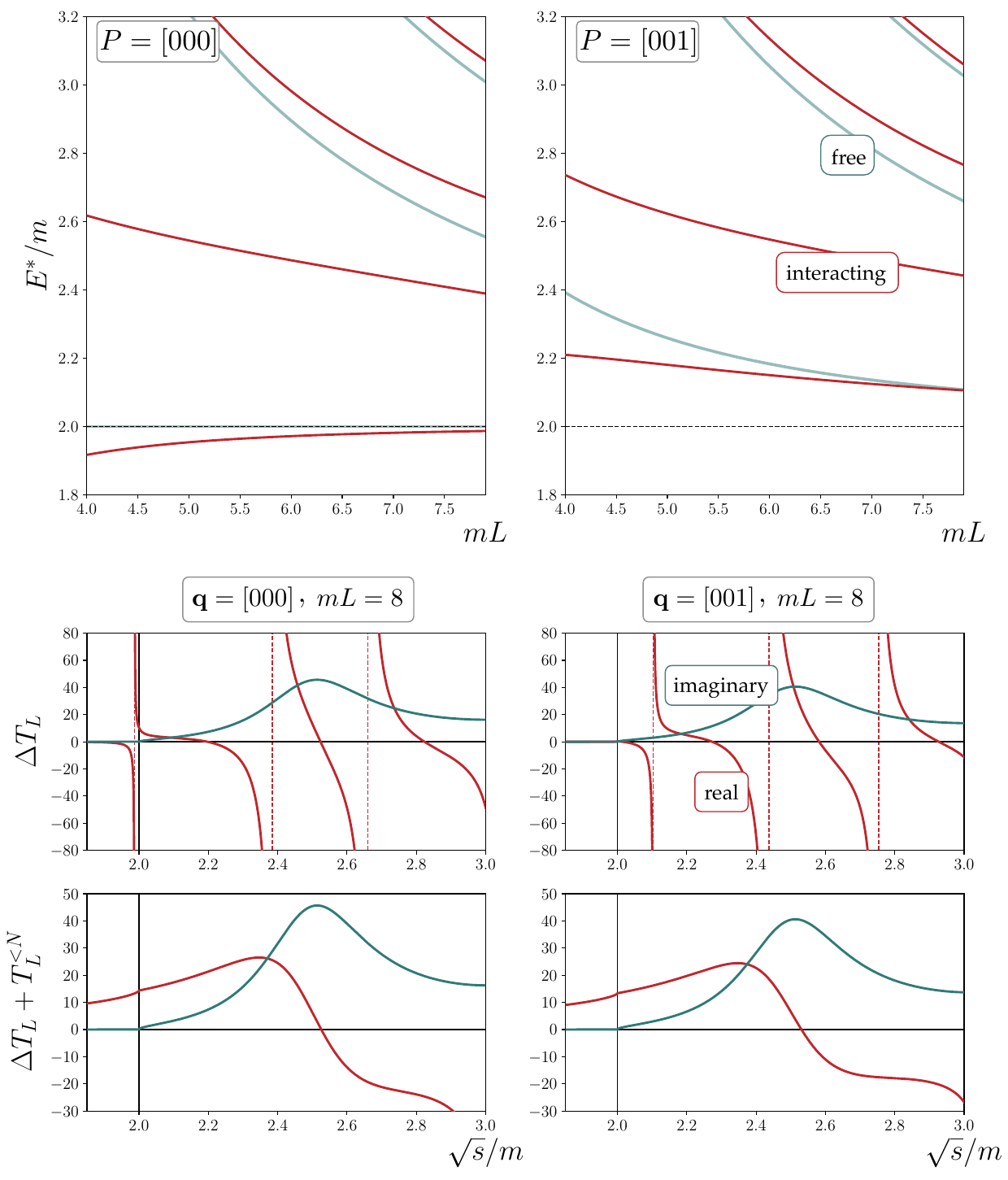}
\caption{[Top panels] Finite-volume spectrum for two values of the spatial momentum. The blue lines show the result for the free theory, while the red lines correspond to the interacting theory with $\mathcal{M}$ given in Fig.~\ref{fig:M_H}.  [Middle panels] $\Delta    T_L$ as defined in Eq.~\eqref{eq:fvcorrection}, given $\mathcal{M}$ and $\mathcal{H}$ in Fig.~\ref{fig:M_H}, where initial and final states are fixed to be at rest. [Bottom panels] The terms defined in the brackets in  Eq.~\eqref{eq:master}. 
}\label{fig:EvsL}
\end{center}
\end{figure}

The truncation of $\mathcal M(s)$ to a single partial wave reduces the two-hadron energy quantization condition, Eq.~\eqref{eq:QC1}, to a simple algebraic relation,
\begin{align}
\mathcal M(s) = -F^{-1}(P,L)    \,,
\label{eq:QCswave}
\end{align}
that can readily be solved numerically.
 The corresponding finite-volume spectrum is shown in the upper panels of Fig.~\ref{fig:EvsL} for two choices of the total three-momentum.

From the unitarity constraint on the $1 \to 2~(2 \to 1)$ transition amplitude, $\mathcal{K}(s)$ must be combined with the real-valued quantity, $\mathbf{H}(s, Q^2)$, as in Eq.~\eqref{eq:Hmat}, to obtain $\mathcal H$. An all-orders perturbation-theory argument requires that for any pole singularity of $\mathcal{K}(s)$, there must be an associated pole singularity in $\mathbf{H}(s, Q^2 )$~\cite{Briceno:2018aml}. This motivates rewriting $\mathbf{H}(s, Q^2 )$ as~\cite{Briceno:2015dca, Briceno:2016kkp}
\begin{align}
\label{eq:H_bold}
\mathbf{H}(s, Q^2) = 
\mathbf{B}(s, Q^2) \mathcal{K}(s) \,,
\end{align}
where $\mathbf{B}(s,Q^2)$ is a smooth real function. Here we take $\mathbf{B}(s,Q^2) = f(Q^2)/3$, where $f(Q^2)$ is the single-particle form factor, which can be parametrized by a simple monopole form,
\begin{align}
f(Q^2)=\frac{1}{1+{Q^2}/m_R^2} \,.
\end{align}
The resulting functional form of $\mathcal{H}(s,Q^2)$ for a range of kinematics is shown in the second panel of Fig.~\ref{fig:M_H}.

For the remainder of the discussion, we consider kinematics for which only the s-channel contributions to $\mathcal T$ are physical. As a result, only these can lead to power-law finite-volume effects, and the $u$-channel contributions can be ignored. Given the specified forms of $\mathcal{M}(s)$ and $\mathcal{H}(s,Q^2)$, one can now determine $ \Delta    T_L$ using Eq.~\eqref{eq:fvcorrection}. The result is plotted in the middle panels of Fig.~\ref{fig:EvsL} for a volume with $m L = 8$, and two values of the total spatial momentum of the current. The initial and final states are fixed to be at rest.  Note that the real part of $\Delta T_L$ exhibits a series of poles that coincide with the energy levels shown in Fig.~\ref{fig:EvsL}. The imaginary piece has a cusp at threshold and has the same qualitative peak structure as $\mathcal{M}(s)$ and $\mathcal{H}(s,Q^2)$. This is consistent with the fact that ${\rm Im } \, \Delta T_L $ satisfies Eq.~\eqref{eq:imDeltaTL}. Furthermore, given Eqs.~\eqref{eq:c} and \eqref{eq:LL_formula1}, $T^{< N}_L$ can be constrained within the parametrization presented. The lower panels of Fig.~\ref{fig:EvsL} plot $T^{< N}_L+\Delta T_L$ for the two spatial momenta considered, exhibiting the exact cancellation of the poles that was emphasized at the end of Sec.~\ref{sec:main_result}.

\begin{figure}[t]
\begin{center}
\includegraphics[width=.9\textwidth]{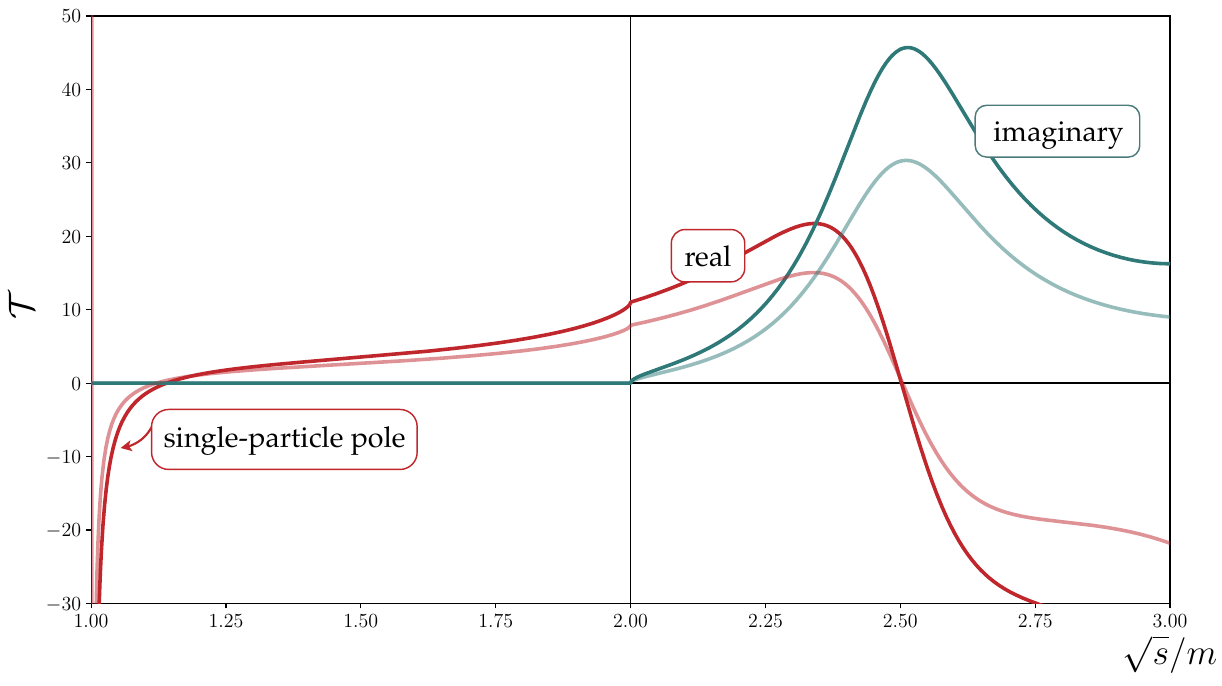}
\caption{Real (red) and imaginary (blue) parts of the Compton amplitude. The initial and final states are fixed to be at rest. The darker lines correspond to currents that have zero momentum, the faded lines correspond to $q=2\pi [001]/L$, where $mL=8$ to match the functions illustrated in Fig.~\ref{fig:EvsL}.}
\label{fig:comp_amp_000}
\end{center}
\end{figure}

Next, we numerically demonstrate that the imaginary part of $\Delta T_L$ is equal to that of $\mathcal{T}$. To proceed, $\mathcal T$ in Eq.~\eqref{eq:compton} needs to be constructed in terms of smooth functions, and for this purpose, the issue of K-matrix poles, mentioned briefly at the end of Sec.~\ref{sec:check}, must be addressed. Note that when Eq.~\eqref{eq:H_bold} is applied, the third term in brackets in Eq.~\eqref{eq:compton} develops real-valued poles corresponding to those of the K matrix,
\begin{align}
i\mathbf{H}(s,Q^2)
\frac{1}{1-i\rho(s) \mathcal{K}(s) }\rho(s) \,
\mathbf{H}(s,Q^2) \ \longrightarrow \ - \mathbf{B}(s,Q^2) \mathcal{K}(s) \mathbf{B}(s,Q^2) \,.
\end{align}
In the full Compton amplitude, this must be exactly canceled by a pole in $\mathbf{T}$, motivating one final re-parametrization 
\begin{align}
\mathbf{T}(s,Q^2) = 
\mathbf{B}(s,Q^2) \, \mathcal{K}(s)\,
\mathbf{B}(s,Q^2) +\mathbf{S}(s,Q^2)  \,,
\end{align} 
where $\mathbf{S}(s,Q^2)$ is a smooth function in the region of interest. We conclude that, in this kinematic region, the analytic structure of the Compton amplitude is given in terms of known functions, plus an additional smooth contribution, $\mathbf{S}(s,Q^2)$, which will be set to vanish for this toy example. This completes our construction of a parametrization for the Compton amplitude, with the result plotted in Fig.~\ref{fig:comp_amp_000}. This numerically confirms the result derived in Sec.~\ref{sec:check} and summarized in Eq.~\eqref{eq:imDeltaTL}, that $\text{Im}\, \Delta T_L = \text{Im} \, \mathcal T$. 

\begin{figure}[t]
\begin{center}
\includegraphics[width=0.9\textwidth]{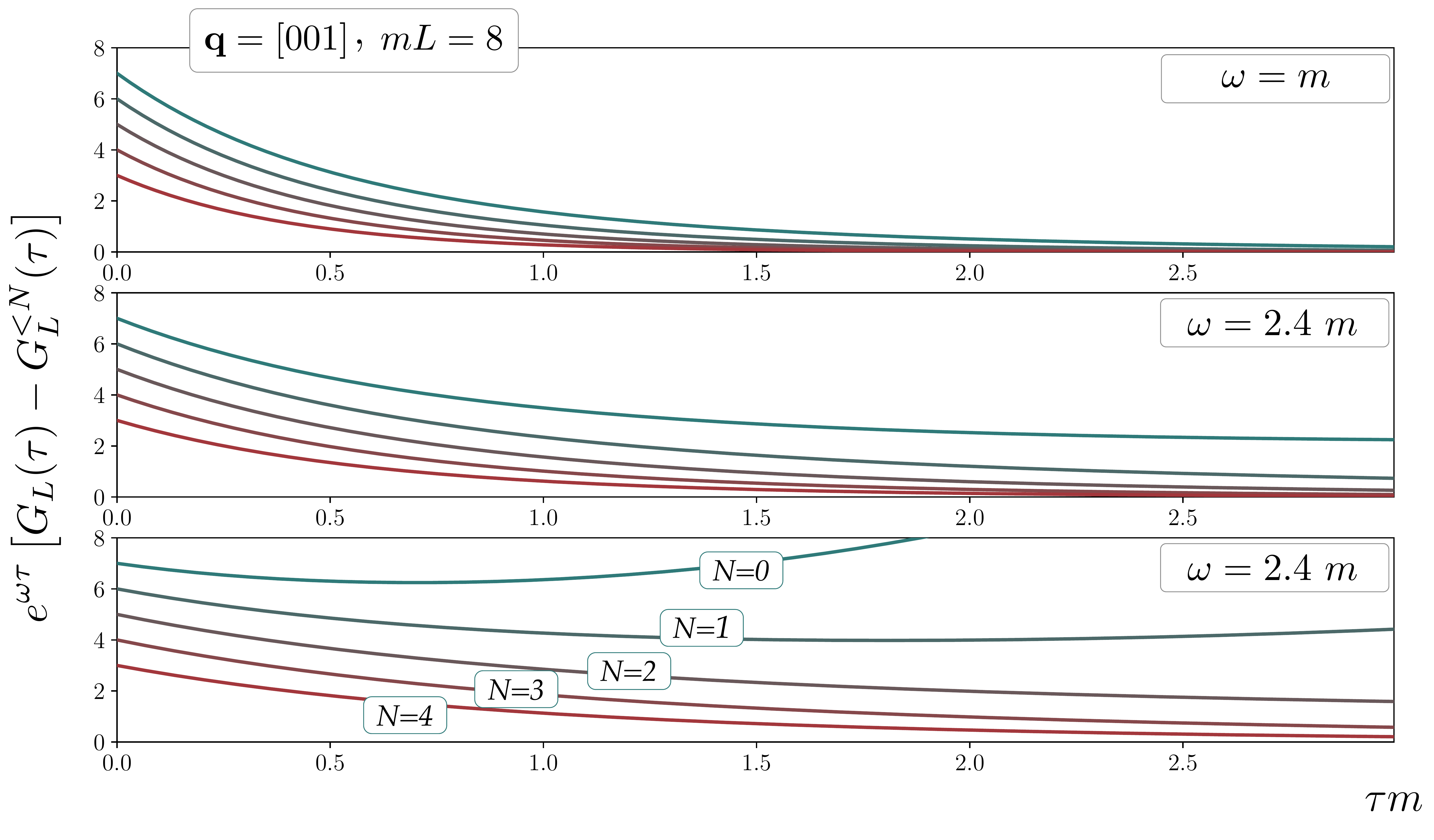}
\caption{ Shown is the integrand of Eq.~\eqref{eq:TgN} for the example of this section for a range of values of $N$ and $\omega$, where initial and final states are fixed to be at rest. The color coding in the upper two panels corresponds to that of the lower panel.  }
\label{fig:Corr_Swave_BW}
\end{center}
\end{figure}

To complete the illustration of the proposed formalism, in Fig.~\ref{fig:Corr_Swave_BW}, we plot the subtracted correlation function, $e^{\omega \tau} \left[ G_L(\tau)-G_L^{<N}(\tau) \right]$, for a set of kinematics and for various subtractions, as a function of Euclidean time. This defines the integrand in Eq.~\eqref{eq:TgN} leading to $T_L^{\geq N}$. While this quantity can be obtained directly from Eq.~\eqref{eq:master} given the constrained values of $\mathcal T$ and $T^{< N}_L+\Delta T_L$ (by performing an inverse integral transform), one can also approximate the function by summing up contributions from a sufficiently large number of lowest intermediate finite-volume states, see Eq.~\eqref{eq:Gdecom}. Nine such states are used to obtain the results of Fig.~\ref{fig:Corr_Swave_BW}. The function exhibits the expected behavior, with an increased number of subtractions accelerating the fall off at large Euclidean separations. It must be stressed that in the case where two-particle states can go on shell, a number of subtractions are required to render the integral convergent. However, in practice it may be profitable to also subtract states in the off-shell region to accelerate the convergence and improve the overall uncertainty on the resulting amplitude. This is demonstrated in the first panel of Fig.~\ref{fig:Corr_Swave_BW}, where the chosen value of $\omega$ does not provide sufficient energy to produce any on-shell intermediate states in this example, but the states close to threshold cause only a slowly damping functions of $\tau$ if not subtracted.

As a final remark, we emphasize that this numerical example only serves as a quantitative demonstration of the various features of the building blocks of the master equation \eqref{eq:master} in a toy model, which, however, was built consistent with physically motivated scenarios. While the approach in this section was to construct the final-volume quantities from the knowledge of infinite-volume amplitudes, in the realistic use of the formalism presented, the known quantities will be the finite-volume two-, three- and four-point Euclidean correlation functions, and the infinite-volume Minkowski observable will subsequently be deduced. In particular, one must note that in the parametrization of this section, the smooth function $\mathbf{S}(s,Q^2)$ is the only unknown infinite-volume quantity whose value can not be fixed by the knowledge of two- and three-point functions in the theory, and only a determination of the four-point correlation function as defined in Eq.~\eqref{eq:GFV} can constrain its value, following the procedure of this work.\footnote{A similar consideration is presented in Refs.~\cite{Shanahan:2017bgi,Tiburzi:2017iux} in the context of matching lattice-QCD results to an effective field theory description of a nuclear double-$\beta$ decay.}

\section{Conclusion and outlook~\label{sec:conclusion}}
\begin{figure}[t!]
\begin{center}
\includegraphics[width=\textwidth]{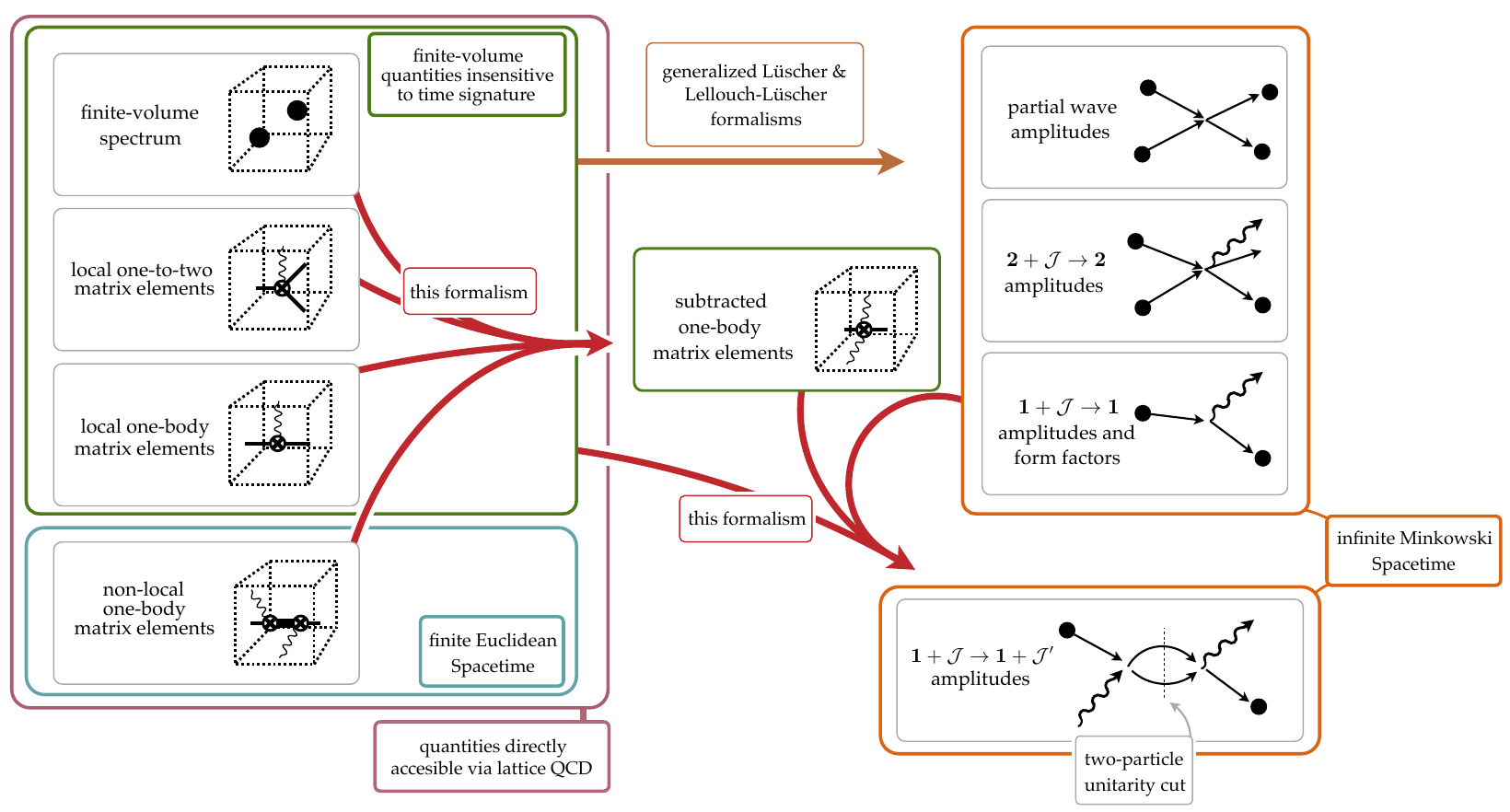}
\caption{Schematic map of the formalism outlined in this work. To extract the infinite-volume Minkowski amplitude involving two currents, a number of building blocks have to be determined with lattice QCD (shown on the left of the figure), including the spectrum and transition amplitudes with single currents. The red arrows show how these building blocks are combined in the current formalism.}
\label{fig:flow_chart}
\end{center}
\end{figure}

\noindent
The formalism presented in this work offers a path from Euclidean finite-volume correlation functions of time-displaced local electroweak currents to long-range contributions to hadronic amplitudes in an infinite Minkowski spacetime. Given the complicated nature of the desired amplitudes, it should not be too surprising that the relation derived in this paper requires a detailed understanding of various building blocks on both ends of the mapping. Figure~\ref{fig:flow_chart} summarizes the conclusions of this work, and provides guidance on how lattice-QCD quantities may be used to access the physical infinite-volume amplitudes of interest.

The main result of this work is summarized in Eq.~\eqref{eq:master}. In short, an extract of the original correlation function is identified that is independent of the time signature in the theory, denoted in Fig.~\ref{fig:flow_chart} as the subtracted one-body matrix element. This is achieved by removing contributions to the finite-volume four-point correlation function that arise from the lowest-lying intermediate states, including all the states that can go on shell. A closed form for the necessary additive piece, $\left[T^{< N}+ \Delta    T\right]_{\mathcal M ,\mathcal H}$ in Eq.~\eqref{eq:master}, is then provided. Not only does this form remove all power-law finite-volume effects below three-hadron productions thresholds, but it also restores the correct analytic structure of the infinite-volume Minkowski amplitude. This additive piece can be evaluated separately from dedicated lattice-QCD studies of two- and three-point functions. 

The approach of this work generalizes the formalism of Refs.~\cite{Christ:2010gi, Christ:2014qaa, Christ:2015pwa}. In particular, the results presented hold not only for single-hadron long-range electroweak transitions, but also for transitions involving the vacuum in the initial/final states, such as for matrix element incurred in studying the QCD structure of the photon~\cite{Ji:2001wha}. Arbitrary quantum numbers, such as spin, flavor, partial waves and total CM momentum, are incorporated in the formalism, and the possibility of multiple coupled partial-wave or flavor channels in intermediate two-hadron states is accounted for. An explicit map of the workflow for future numerical implementation of the formalism is shown in Fig.~\ref{fig:eqsflowchart}, with a reference to quantities that are defined in various equations throughout this paper.

The general framework of this work can further serve as guidance on how to address more complex scenarios, such as considering two-hadron initial and final states (relevant for neutrinoful and neutrinoless double-$\beta$ decay), or extending the kinematic reach of the problem so that more than two hadrons can be produced on shell in the intermediate states. Extension to kinematic regions beyond the three-hadron productions may be possible given the recently-developed technologies for the determination of three-hadron observables from a finite-volume study~\cite{Polejaeva:2012ut,Briceno:2012rv, Hansen:2014eka,Hansen:2015zga, Hammer:2017uqm,Hammer:2017kms, Guo:2017ism,Mai:2017bge, Briceno:2017tce, Doring:2018xxx, Briceno:2018mlh, Mai:2018djl, Briceno:2018aml, Blanton:2019igq,Hansen:2019nir}. More immediately, one can imagine extending these ideas for processes like $\gamma^*\gamma^* \to \pi\pi$, which would be relevant for dispersive analyses of the hadronic light-by-light contribution to muon $g-2$~\cite{Colangelo:2015ama}.

 \section{Acknowledgements}
RAB is supported in part by USDOE grant No. DE-AC05-06OR23177, 
under which Jefferson Science Associates, LLC, manages and operates Jefferson Lab.
RAB also acknowledges support from the USDOE Early Career award, contract de-sc0019229. ZD is supported by the Alfred P. Sloan Foundation, and by the Maryland Center for Fundamental Physics at the University of Maryland, College Park. 
AB and MRS acknowledge support from the U.S. Department of Energy, Office of Science,
Office of Nuclear Physics, under Award Number DE-SC0019647.
RAB would like to thank J. Dudek and the rest of the Hadron Spectrum Collaboration for useful conversations. 

\bibliography{bibi} 
\end{document}